\documentclass[sigconf]{acmart}

\usepackage{multirow}
\usepackage{tabularx}
\usepackage{colortbl}
\usepackage{enumitem}

\definecolor{Gray}{gray}{0.9}

\copyrightyear{2021}
\acmYear{2021}
\setcopyright{acmlicensed}\acmConference[ACSAC '21]{Annual Computer Security Applications Conference}{December 6--10, 2021}{Virtual Event, USA}
\acmBooktitle{Annual Computer Security Applications Conference (ACSAC '21), December 6--10, 2021, Virtual Event, USA}
\acmPrice{15.00}
\acmDOI{10.1145/3485832.3488020}
\acmISBN{978-1-4503-8579-4/21/12}

\acmBadgeR[https://www.acm.org/publications/policies/artifact-review-and-badging-current]{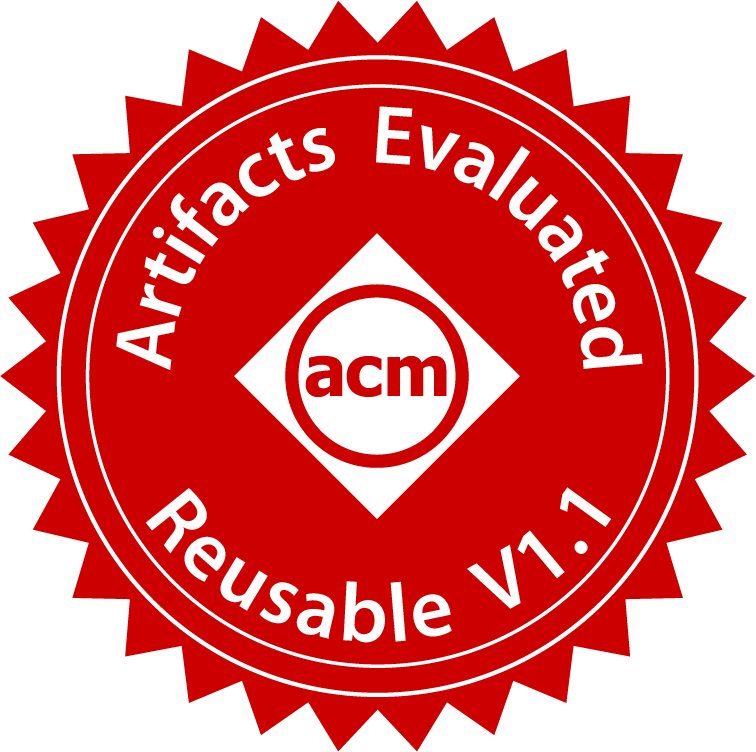}

\begin{document}

\title[Reproducible and Adaptable Log Data Generation for Sound Cybersecurity Experiments]{Reproducible and Adaptable\\ Log Data Generation for\\ Sound Cybersecurity Experiments}

\author{Rafael Uetz}
\affiliation{%
  \institution{Fraunhofer FKIE}
  \streetaddress{Zanderstraße 5}
  \postcode{53177}
  \city{Bonn}
  \country{Germany}}
\email{rafael.uetz@fkie.fraunhofer.de}
\author{Christian Hemminghaus}
\affiliation{%
  \institution{Fraunhofer FKIE}
  \streetaddress{Zanderstraße 5}
  \postcode{53177}
  \city{Bonn}
  \country{Germany}}
\email{c.hemminghaus@fkie.fraunhofer.de}
\author{Louis Hackländer}
\affiliation{%
  \institution{Fraunhofer FKIE}
  \streetaddress{Zanderstraße 5}
  \postcode{53177}
  \city{Bonn}
  \country{Germany}}
\email{louis.hacklaender@fkie.fraunhofer.de}
\author{Philipp Schlipper}
\affiliation{%
  \institution{Fraunhofer FKIE}
  \streetaddress{Zanderstraße 5}
  \postcode{53177}
  \city{Bonn}
  \country{Germany}}
\email{philipp.schlipper@fkie.fraunhofer.de}
\author{Martin Henze}
\affiliation{%
  \institution{RWTH Aachen University}
  \streetaddress{Ahornstraße 55}
  \postcode{52074}
  \city{Aachen}
  \country{Germany}}
\affiliation{%
  \institution{Fraunhofer FKIE}
  \streetaddress{Zanderstraße 5}
  \postcode{53177}
  \city{Bonn}
  \country{Germany}}
\email{henze@cs.rwth-aachen.de}

\begin{abstract}
Artifacts such as log data and network traffic are fundamental for cybersecurity research, e.g., in the area of intrusion detection.
Yet, most research is based on artifacts that are not available to others or cannot be adapted to own purposes, thus making it difficult to reproduce and build on existing work.
In this paper, we identify the challenges of artifact generation with the goal of conducting sound experiments that are valid, controlled, and reproducible.
We argue that testbeds for artifact generation have to be designed specifically with reproducibility and adaptability in mind.
To achieve this goal, we present SOCBED, our proof-of-concept implementation and the first testbed with a focus on generating realistic log data for cybersecurity experiments in a reproducible and adaptable manner.
SOCBED enables researchers to reproduce testbed instances on commodity computers, adapt them according to own requirements, and verify their correct functionality.
We evaluate SOCBED with an exemplary, practical experiment on detecting a multi-step intrusion of an enterprise network and show that the resulting experiment is indeed valid, controlled, and reproducible.
Both SOCBED and the log dataset underlying our evaluation are freely available.
\end{abstract}

\ccsdesc[500]{Security and privacy~Intrusion/anomaly detection and malware mitigation}
\ccsdesc[300]{Security and privacy~Network security}
\ccsdesc[300]{Computing methodologies~Modeling and simulation}

\begin{CCSXML}
  <ccs2012>
    <concept>
      <concept_id>10002978.10002997</concept_id>
      <concept_desc>Security and privacy~Intrusion/anomaly detection and malware mitigation</concept_desc>
      <concept_significance>500</concept_significance>
      </concept>
    <concept>
      <concept_id>10002978.10003014</concept_id>
      <concept_desc>Security and privacy~Network security</concept_desc>
      <concept_significance>300</concept_significance>
      </concept>
    <concept>
      <concept_id>10010147.10010341</concept_id>
      <concept_desc>Computing methodologies~Modeling and simulation</concept_desc>
      <concept_significance>300</concept_significance>
      </concept>
  </ccs2012>
\end{CCSXML}

\keywords{log data, testbed, reproducibility, intrusion detection, cybersecurity}

\maketitle

\section{Introduction}
\label{sec:introduction}

Successful cyberattacks against organizations' computer networks have ramped up in quantity and severity over the past years~\cite{verizon2021data}.
As a recent example, the 2020 SolarWinds hack alone affected thousands of companies and United States government offices~\cite{reuters2021solarwinds}.
Timely detecting such breaches and thus stopping adversaries before they reach their final goals requires indicators of adversary activity.
\emph{Log data} provide numerous valuable sources of such indicators, ranging from operating system logs (e.g., Windows Event Logs or syslogs) over service logs (e.g., Apache's Common Logs) to dedicated security system 
alerts (e.g., from firewalls, intrusion detection systems (IDSs), or endpoint protection agents).
Due to this large number of different log data sources, thoughtful configuration and analysis of these sources is vital for intrusion detection~\cite{barse2004extracting,verizon2011data,yen2013beehive}.

To aid in this task, companies employ \emph{security information and event management} (SIEM) systems, which try to tackle the task of intrusion detection with several rule-based and anomaly-based methods~\cite{bhatt2014operational}, but are far from being perfect~\cite{maisey2014moving}.
Consequently, current research is concerned with questions such as how events or alerts can be enriched, prioritized, or correlated~\cite{roundy2017smoke,najafi2019malrank,veeramachaneni2016ai2} as well as how adversaries can be modeled to improve the discovery of cyberattacks~\cite{microsoft2021bayes}.

Any research dealing with these questions must be backed by sound evaluations -- which require meaningful log data to evaluate against.
Unfortunately, there is a significant lack of such data in the scientific community~\cite{turcotte2019unified} and freely available datasets usually do not match researchers' requirements for novel experiments~\cite{sharafaldin2018toward}.

Consequently, a recent survey by Landauer et al.~\cite{landauer2020system}, e.g., found that almost 60\,\% of papers in the field of log clustering rely on unpublished datasets for evaluation and the majority of those using public datasets concentrates on only two of them.
While unpublished datasets prevent reproduction of findings, published yet fixed datasets are limited in scope, not adaptable (e.g., w.r.t.\ changes and up-to-dateness), and their creation process might lack transparency~\cite{vandewalle2009reproducible}.
As such, existing fixed datasets are often pointless for novel experiments, as unchangeable underlying scenarios or system configurations do not match the requirements of these experiments.

In this paper, we study the question of how to generate meaningful, reproducible, and adaptable log datasets for sound scientific cybersecurity experiments to address the lack of suitable and freely available datasets that can be adapted to the requirements of novel experiments.
By formulating design goals for sound experiments in log data research, we find a need for dedicated and publicly available testbeds to efficiently generate suitable and realistic log datasets as they would arise in a real enterprise network in an adaptable and reproducible manner.
To address this need, we present and evaluate SOCBED, a proof-of-concept testbed allowing for a reproducible and adaptable generation of log datasets. 
SOCBED enables researchers to better build on existing work by reusing existing scenarios and consequently save the effort of building own testbeds from scratch while at the same time improving the comparability of results.

We present the following contributions:
\begin{itemize}[leftmargin=*] % COMMENT: better bullet alignment -> [leftmargin=6.6mm]
\item We survey the field of log data generation for cybersecurity experiments and find that data collection in productive networks or proprietary testbeds leads to experiments that often lack validity, controllability, and reproducibility.
\item To remedy this situation, we derive design goals for sound experiments in cybersecurity research, specifically focusing on the generation of realistic, transparent, adaptable, replicable, and available artifacts such as log datasets.
\item To showcase and validate our approach, we present SOCBED, a self-contained open-source cyberattack experimentation testbed with a focus on generating reproducible and adaptable log datasets, e.g., for intrusion detection research.
SOCBED simulates a company network with clients, servers, and common services as well as benign user activity and an adversary performing multi-step attacks.
The testbed can be built and run on a commodity PC and is freely available~\cite{uetz2021socbed}.
\item We use SOCBED to perform a practical attack detection experiment and show that this experiment is reproducible on commodity PCs, yields meaningful results, and allows for an adaptation of log data generation in a controlled manner.
The generated dataset is also publicly available~\cite{uetz2021dataset}.
\end{itemize}

The remainder of this paper is structured as follows.
In Section~\ref{sec:problemanalysis} we formulate challenges of acquiring log data for cybersecurity research and motivate the need for reproducible and adaptable log datasets.
Subsequently, we derive design goals for sound cybersecurity experiments in Section~\ref{sec:design} and analyze to which extent related work meets these goals in Section~\ref{sec:relatedwork}.
To fill the gap of a testbed particularly targeting the generation of reproducible and adaptable log data, we present SOCBED in Section~\ref{sec:implementation}.
We evaluate the reproducibility and adaptability of SOCBED by performing an exemplary experiment in Section~\ref{sec:evaluation}, before discussing SOCBED's design decisions and resulting limitations in Section~\ref{sec:discussion}.
Section~\ref{sec:conclusion} concludes this paper.

\section{Log Data in Cybersecurity Research}
\label{sec:problemanalysis}

Log data are indispensable and extremely valuable sources for the timely detection of network breaches~\cite{verizon2011data}.
Consequently, they provide the foundation for various streams of research, e.g., w.r.t.\ enrichment, prioritization, and correlation of events~\cite{roundy2017smoke,najafi2019malrank,veeramachaneni2016ai2} or the realistic modeling of adversaries~\cite{microsoft2021bayes}.
However, although being required as foundation for sound evaluations, there is a significant lack of meaningful log data in the scientific community~\cite{sharafaldin2018toward,bowen2016enabling,turcotte2019unified}.
In the following, we discuss why collecting sound log data from productive networks is difficult and why fixed datasets generated from proprietary testbeds have several drawbacks as well (Section~\ref{sec:background:acquiring-log-data}).
Subsequently, we argue that the resulting limitations are an obstacle for the reproducibility of log data research and for building on existing work (Section~\ref{sec:background:missing-reproducibility}) and argue how adaptable log datasets can remedy this situation (Section~\ref{sec:background:motivation}).

\subsection{Challenges of Acquiring Log Data}
\label{sec:background:acquiring-log-data}

Log data as required for intrusion detection research are usually generated by assets as they are typically found in company networks, i.e., operating systems, services, and dedicated security products such as firewalls, network-based intrusion detection systems (NIDSs), and endpoint protection agents.
Depending on the desired experiment, logs of benign user activity and/or realistic cyberattacks are required.
To achieve this goal, log data acquisition can be done in two fundamentally different ways:
Collection in a productive network with real users or generation in a dedicated, controlled lab environment.
Both sources come with specific advantages and disadvantages, which are discussed in the following.

While collecting real-world log data in a productive network has the obvious advantage of providing realistic data, it also comes with significant drawbacks:
Most importantly, the variety of successful cyberattacks may be too small for meaningful evaluations because the productive network is either not vulnerable to the attacks, their implementation is too costly, or attack execution is deemed too dangerous and thus not permitted.
Likewise, confidentiality or privacy issues often forbid the publication of collected data or necessitate extensive anonymization, severely reducing utility for other researchers~\cite{turcotte2019unified}.
Furthermore, as there is only one instance of each productive network and its state always changes, collected data are neither replicable at a later point in time nor reproducible by other researchers.
This leads to a lack of controllability: It is not possible to examine the effect of a changed parameter that affects log data generation (e.g., configuration change) in an isolated way.
Finally, the adaptability of the productive network is usually limited. 
It might not be possible to add, remove, or exchange certain assets as required for an evaluation.
In particular, other researchers without access to the network cannot perform configuration changes that might be required for subsequent experiments.

Consequently, researchers often rely on dedicated lab testbeds for log data acquisition to avoid these issues~\cite{davis2013survey}.
Log datasets generated by such testbeds are usually not affected by confidentiality or privacy concerns and can therefore be made available.
However, typically only datasets are published, but the testbeds with which they were generated are not~\cite{moustafa2015unsw,bowen2016enabling,sharafaldin2018toward}.
As all datasets are created with a specific use case in mind (e.g., IDS evaluation), they often do not fit the requirements of other researchers even though the underlying testbed could be adapted to generate the desired data if it was publicly available.
Thus, researchers who require slightly different data often have to create an own testbed instead of using an existing dataset.

\subsection{Missing Reproducibility and Adaptability}
\label{sec:background:missing-reproducibility}

Although being one of the most important properties of scientific experiments~\cite{peisert2007design}, reproducibility is often limited in current log data research~\cite{landauer2020system}.
Besides being an integral part of sound experiments, reproducibility also facilitates adaptability and thus spurs further research:
With the data and information required to reproduce others' findings, experiments can be (slightly) adapted to study novel research questions. 
Specifically focusing on the research area of intrusion detection, numerous works~\cite{khraisat19survey} present and use once generated datasets without the ability to reproduce or adapt them, let alone the option to adjust them to other use cases.
To illustrate this issue, Sharafaldin et al. \cite{sharafaldin2018toward} provide an overview over publicly available datasets ranging from network packet to system call captures generated for IDS training.
For all of these datasets, the testbeds used to generate the dataset are not made available, thus preventing reproduction or adaptation.

Even worse, for a large batch of work on log data and intrusion detection, the underlying log data are not (publicly) available \emph{at all}, rendering the reproduction (and thus also extension) of their experiments impossible.
Examples include MalRank~\cite{najafi2019malrank}, Smoke Detector~\cite{roundy2017smoke}, and Beehive~\cite{yen2013beehive}.
We assume that the main reason for not disclosing these datasets are confidentiality or privacy concerns.

\subsection{The Case for Adaptable Log Datasets}
\label{sec:background:motivation}

Given the missing reproducibility and adaptability of existing log datasets, using these as a basis for novel experiments is often meaningless because the datasets' underlying scenarios or configurations differ from what is required for novel experiments.
Common issues include outdated scenario components (e.g., obsolete operating systems or attacks that are no longer prevalent in the real world), a scenario not matching the new experiment's context (e.g., Windows vs.\ Linux clients/servers or different security measures in place), or a logging configuration not producing the logs required for the evaluation (e.g., logs produced by an up-to-date Sysmon version are required as input for an IDS).
We encountered such issues in several experiments in the context of intrusion detection, leaving us with no other choice than building testbeds from scratch to generate log data instead of using existing log datasets.
Given these problems with missing reproducibility and adaptability of log data research, especially in the context of intrusion detection, we set out to remedy this situation with our contributions in this paper.

To achieve this goal, we are convinced that long-ranging, usable log datasets for sound cybersecurity research must be subject to frequent updates and modifications by different groups of researchers.
This can only be achieved by an open-source testbed specifically built for easy reproducibility and adaptability, thus allowing a large number of researchers to reproduce log datasets and adapt, e.g., the logging configuration, while retaining the same scenario.
Vice versa, the scenario (e.g., systems, services, or attacks) can be adapted or extended while still producing the same types of log data.

\section{Design Goals for Sound Cybersecurity Experiments}
\label{sec:design}

\begin{figure}
  \centering
  \includegraphics[width=\linewidth]{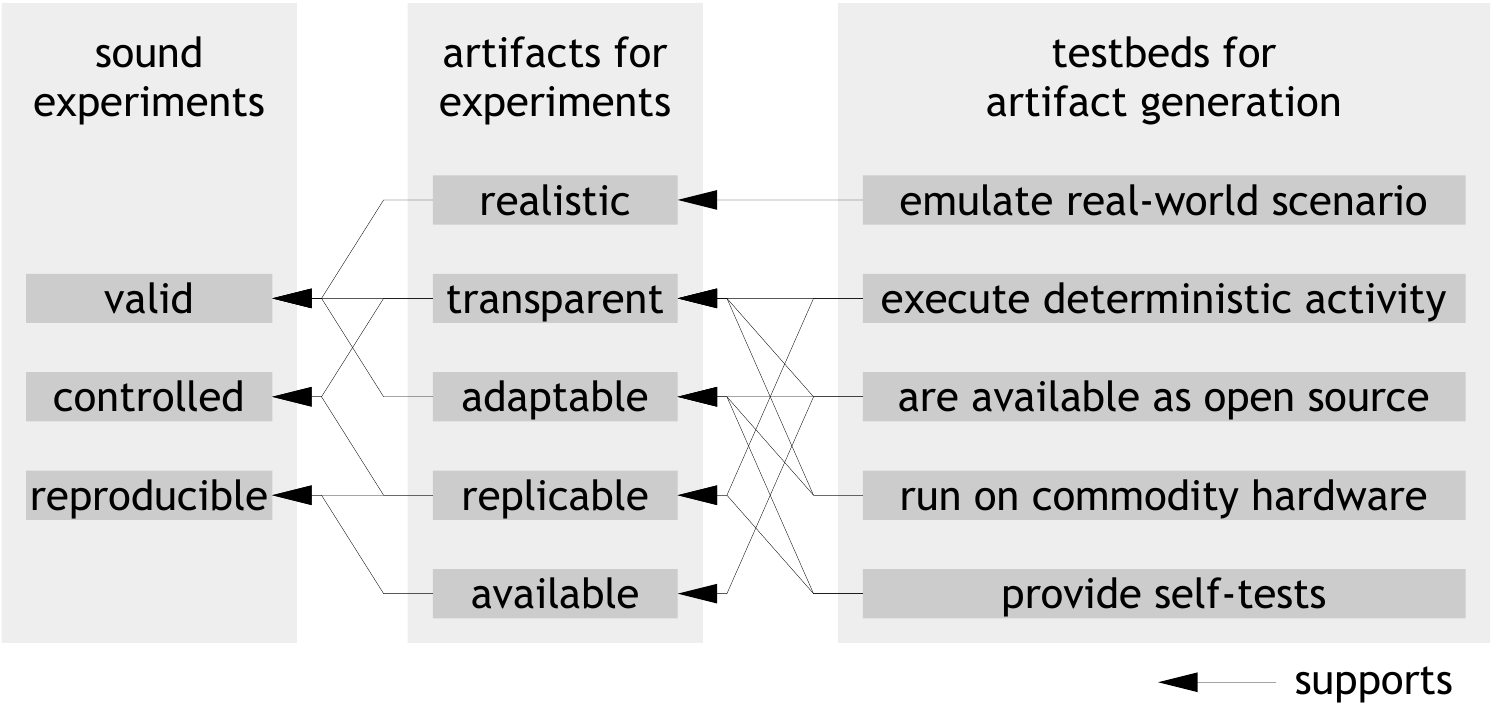}
  \caption{Conducting sound experiments imposes requirements on the used artifacts (e.g., log data), which in turn impose requirements on testbeds used for their generation.}
  \label{fig:requirements}
  \Description{Block diagram summarizing the properties of artifacts and testbeds that support sound experiments, as described in the text.}
\end{figure}

Our analysis of the use of log data in cybersecurity research identifies several pitfalls when performing experiments based on log data, especially w.r.t.\ reproducibility and adaptability.
In the following, we summarize these issues and derive design goals for sound cybersecurity experiments.
In this process, we abstract from log data to artifacts in general, but still focus on log data in the examples given.
As summarized in Figure~\ref{fig:requirements}, we start by discussing three vital properties of \emph{scientific experiments} in general: validity, controllability, and reproducibility (Section~\ref{sec:reqsci}).
From these properties, we derive design goals for \emph{artifacts used by such experiments} (Section~\ref{sec:reqart}).
Finally, we derive design goals for \emph{testbeds for artifact generation} (Section~\ref{sec:reqgen}).
As a result, we argue that sound experiments that require artifacts such as log data benefit strongly from testbeds that allow for (1) realistic scenarios and (2) deterministic activity while being easy to use for other researchers because they are (3) open source, (4) can be run on commodity hardware, and (5) provide self-tests to verify correct functionality after installation or adaptation.

\subsection{Properties of Sound Experiments}
\label{sec:reqsci}

Cybersecurity is both an engineering discipline and a science~\cite{peisert2007design}.
In both fields, \emph{sound experiments} are fundamentally important to test hypotheses and thus advance knowledge~\cite{hepburn2021scientific, montgomery2017design}.
While there is no generally accepted definition of soundness~\cite{hepburn2021scientific}, literature on the design of experiments often states that a sound experiment must allow for \emph{valid} conclusions and be \emph{controlled} and \emph{reproducible}~\cite{peisert2007design,montgomery2017design,blackburn2016truth}.
We now describe why these properties are important in general and for log data-based experiments in particular.

\paragraph{Valid}
Validity is the extent to which a concept, conclusion, or measurement is well-founded and rules out alternative explanations~\cite{rich2018empirical}.
We can distinguish \emph{internal validity} (i.e., the confidence in the conclusions drawn in the strict context of an experiment) from \emph{external validity} (i.e., the degree to which conclusions can be generalized and/or applied to the real world)~\cite{mcdermott2011internal}.
In the field of cybersecurity, there is an ongoing debate on the external validity of research that is based on inappropriate datasets such as the old and overused DARPA/KDD'99 datasets~\cite{sommer2010outside,sharafaldin2018toward,moustafa2015unsw}.

\paragraph{Controlled}
Each experiment can be seen as a process that transforms some input (e.g., attacks against a computer network) into some output (e.g., intrusion alerts)~\cite{montgomery2017design}.
There are usually multiple  variables that affect the outcome, some of which can be controlled by the experimenter (e.g., timing of the attacks) while others can not (e.g., activity of background processes).
Controllability is the extent to which variables can be controlled.
Ideally, an experiment can be repeated multiple times while changing exactly one variable at a time to reliably examine cause-effect relationships~\cite{peisert2007design}.

\paragraph{Reproducible}
An experiment is called reproducible if its results can be reproduced by other researchers~\cite{bajpai2019reproducibility}.
This serves multiple purposes:
(1) Results become more trustworthy when they are independently verified by other researchers~\cite{montgomery2017design},
(2) building on existing research and advancing the state of the art often requires reproduction of existing results in the first place~\cite{bajpai2019reproducibility}, and
(3) building on original experiments also makes results directly comparable.

\subsection{Research Artifacts for Sound Experiments}
\label{sec:reqart}

To avoid the previously discussed pitfalls, we now describe five properties that \emph{artifacts used in experiments} should exhibit to facilitate sound experiments (cf.\ Figure~\ref{fig:requirements}).

\paragraph{Realistic}
For experiment results to be transferable to real-world use cases and thus be externally valid, the used artifacts must resemble key properties of real-world data~\cite{sommer2010outside}.
The notion of realism depends strongly on the context of a concrete experiment.
For example, most IDSs only consider the contents of network packets, not their timing, so a dataset with realistic contents but unrealistic timing would still be valid.
On the contrary, some IDSs might consider timing, making the same dataset invalid for their evaluation.

\paragraph{Transparent}
All relevant details on the contents of artifacts should be made transparent to other researchers. A failure to do so can lead to incorrect assumptions about capabilities and limitations of a dataset and ultimately to invalid conclusions~\cite{bajpai2019reproducibility}.
Non-transparent artifacts can also lead to uncontrolled experimental behavior in case changed variables cause unexpected side effects.

\paragraph{Adaptable}
Adaptability is the extent to which artifacts such as log data can be recreated with changed parameters (e.g., updated cyberattacks).
Adaptability supports validity because it allows experimenters to adapt artifacts depending on the needs of a new experimental context (e.g., re-running an experiment with an attacked system updated from Windows 8 to Windows 10 to regain external validity for real-world application).

\paragraph{Replicable}
Replicability is the extent to which artifacts can be recreated under the same conditions~\cite{bajpai2019reproducibility}, e.g., running the same testbed on the same host.
It is a prerequisite for controlled experiments as it allows for multiple iterations with changed parameters to analyze cause-effect relationships~\cite{montgomery2017design}.
It is also a prerequisite for reproducibility, which allows other researchers to obtain similar results under different conditions (e.g., same testbed, different host).

\paragraph{Available}
Lastly, for an experiment to be reproducible by other researchers, the used artifacts must be made (freely) available to them, either as a dataset or as a testbed that allows to reproduce the artifacts on own hardware.

\subsection{Testbeds for Artifact Generation}
\label{sec:reqgen}

In Section~\ref{sec:problemanalysis}, we argued that artifacts collected from productive systems or networks usually do not allow for controlled and reproducible experiments and thus, dedicated testbeds are an essential means to create artifacts for sound experiments.
In the following, we present five principal properties that a \emph{testbed for artifact generation} should possess to fulfill the previously described properties of artifacts for sound experiments (cf.\ Figure~\ref{fig:requirements}).

\paragraph{Real-World Scenario}
Carefully recreating a real-world scenario in a testbed is vital for the generation of realistic artifacts~\cite{sharafaldin2018toward}.
For example, experiments analyzing intrusions of enterprise networks require a testbed scenario with realistic topology (network zones etc.), assets (operating systems, services, etc.), and activity (benign user activity, cyberattacks, etc.).

\paragraph{Deterministic Activity}
Any activity in the testbed should be performed in a deterministic way to ensure transparent and replicable artifacts.
For example, attacks should be scripted instead of performed manually.
If stochastic activity is required (e.g., to train anomaly detection systems), it should be pseudo-random with a configurable seed, thus making it replicable.

\paragraph{Open Source}
Providing a testbed as open-source software has multiple advantages:
(1) Artifact generation becomes transparent for other researchers, enabling a detailed analysis of why/how certain artifacts are created,
(2) artifact generation becomes adaptable for others so that they can build on previous work, and
(3) log data generated by open-source testbeds are usually not affected by privacy or confidentiality concerns, improving their availability.

\paragraph{Commodity Hardware}
Some testbeds build on multiple physical systems and/or specialized hardware such as proprietary traffic generators~\cite{ferguson2014national,chadha2016cybervan}, making them costly to reproduce for other researchers.
It is therefore beneficial if a testbed can be run on commodity hardware (i.e., common desktop, notebook, or server computers) to aid transparency (because others can re-run a scenario and better understand the generated artifacts) and adaptability (because adapted versions can be run on own hardware).

\paragraph{Self-Tests}
Testbeds are complex and consist of multiple interdependent components, making their installation and operation prone to errors.
A testbed should therefore provide self-tests to verify that all components function correctly after installation.
Self-tests improve replicability of artifacts because potential errors can be found and fixed.
They also aid adaptability because changes that break existing functionality can be identified~\cite{bajpai2019reproducibility}.

\medskip We would like to note that a complete fulfillment of all desired properties might not be possible.
Some goals might even contradict each other, e.g., perfect replicability can impede realism because Internet connectivity must be disabled to avoid non-deterministic network traffic such as software updates.
It is thus the duty of an experiment designer to find appropriate trade-offs during artifact generation to facilitate sound experiments.

In the following section, we analyze existing testbeds and show that all of them have major flaws with regard to the presented requirements, thus impeding their utility for the generation of log data for sound cybersecurity experiments.

\section{Analysis of Related Work}
\label{sec:relatedwork}

Different streams of related work address the challenge of generating meaningful and adaptable artifacts for cybersecurity experiments.
As artifacts collected from productive systems or networks typically cannot be made publicly available due to confidentiality or privacy concerns and are inherently non-replicable (cf.\ Section~\ref{sec:problemanalysis}), dedicated testbeds are the predominant approach to create such artifacts.
Testbeds for artifact generation can be classified into three categories~\cite{davis2013survey}:
An (1) \emph{overlay} simulates or emulates desired functionality (e.g., cyberattacks) on top of an existing, usually productive network, a (2) \emph{simulation} employs an abstracted model instead of real networks or machines, and an (3) \emph{emulation} makes use of full-featured (i.e., virtualized or physical) systems.

Considering our requirements for testbeds for artifact generation underlying sound cybersecurity experiments (cf.\ Section~\ref{sec:reqgen}), we require testbeds to be implemented with deterministic activity.
Thus, overlay testbeds, which are realized on top of uncontrollable networks do not fit our requirements.
Likewise, simulations are not well-suited for generating sound artifacts, as, due to abstraction, they cannot generate realistic log data in the same way as complex, real software such as operating systems.
Consequently, we focus our analysis of related work on testbeds relying on emulation.
We provide an overview of our analysis in Table \ref{tab:relatedwork}.

\begin{table}
  \caption{Our analysis of existing testbeds from related work identifies substantial gaps in meeting the requirements for sound log data generation.}
  \small
  \centering
  \setlength\tabcolsep{1.55mm}
  \begin{tabularx}{\linewidth}{Xp{15pt}p{15pt}p{15pt}p{15pt}p{30pt}}
    \toprule
    \textbf{Testbed name or author} &
      \rotatebox{30}{Real-World Scenario} &
      \rotatebox{30}{Deterministic} &
      \rotatebox{30}{Open Source} &
      \rotatebox{30}{Commodity HW} &
      \rotatebox{30}{Self-Tests} \\
    \midrule
    LARIAT~\cite{rossey2002lincoln} &
      \includegraphics{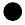} &
      \includegraphics{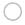} &
      \includegraphics{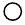} &
      \includegraphics{circle-empty.pdf} &
      \includegraphics{circle-empty.pdf} \\
    National Cyber Range~\cite{ferguson2014national} &
      \includegraphics{circle-full.pdf} &
      \includegraphics{circle-gray.pdf} &
      \includegraphics{circle-empty.pdf} &
      \includegraphics{circle-empty.pdf} &
      \includegraphics{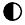} \\
    CyberVAN~\cite{chadha2016cybervan} &
      \includegraphics{circle-half.pdf} &
      \includegraphics{circle-empty.pdf} &
      \includegraphics{circle-half.pdf} &
      \includegraphics{circle-half.pdf} &
      \includegraphics{circle-empty.pdf} \\
    ViSe~\cite{richmond2005vise} &
      \includegraphics{circle-half.pdf} &
      \includegraphics{circle-half.pdf} &
      \includegraphics{circle-half.pdf} &
      \includegraphics{circle-half.pdf} &
      \includegraphics{circle-empty.pdf} \\
    DETERLab~\cite{benzel2011science} &
      \includegraphics{circle-full.pdf} &
      \includegraphics{circle-gray.pdf} &
      \includegraphics{circle-half.pdf} &
      \includegraphics{circle-empty.pdf} &
      \includegraphics{circle-empty.pdf} \\
    ATT\&CK Evaluations~\cite{mitre2020attackevals} &
      \includegraphics{circle-half.pdf} &
      \includegraphics{circle-empty.pdf} &
      \includegraphics{circle-half.pdf} &
      \includegraphics{circle-empty.pdf} &
      \includegraphics{circle-empty.pdf} \\
    DetectionLab~\cite{long2021detection} &
      \includegraphics{circle-half.pdf} &
      \includegraphics{circle-empty.pdf} &
      \includegraphics{circle-full.pdf} &
      \includegraphics{circle-full.pdf} &
      \includegraphics{circle-half.pdf} \\
    SimuLand~\cite{microsoft2021simuland} &
      \includegraphics{circle-half.pdf} &
      \includegraphics{circle-empty.pdf} &
      \includegraphics{circle-half.pdf} &
      \includegraphics{circle-empty.pdf} &
      \includegraphics{circle-empty.pdf} \\
    Skopik et al.~\cite{skopik2014semi} &
      \includegraphics{circle-full.pdf} &
      \includegraphics{circle-half.pdf} &
      \includegraphics{circle-empty.pdf} &
      \includegraphics{circle-full.pdf} &
      \includegraphics{circle-empty.pdf} \\
    Landauer et al.~\cite{landauer2021have} &
      \includegraphics{circle-full.pdf} &
      \includegraphics{circle-half.pdf} &
      \includegraphics{circle-empty.pdf} &
      \includegraphics{circle-full.pdf} &
      \includegraphics{circle-empty.pdf} \\
    \midrule
    SOCBED (this paper) &
      \includegraphics{circle-full.pdf} &
      \includegraphics{circle-half.pdf} &
      \includegraphics{circle-full.pdf} &
      \includegraphics{circle-full.pdf} &
      \includegraphics{circle-full.pdf} \\
    \bottomrule
    \multicolumn{6}{c}{} \\
    \multicolumn{6}{c}{
      Requirement fulfilled:
      \enspace\includegraphics{circle-full.pdf} yes
      \enspace\includegraphics{circle-half.pdf} partially
      \enspace\includegraphics{circle-empty.pdf} no
      \enspace\includegraphics{circle-gray.pdf} not discussed} \\
  \end{tabularx}
  \label{tab:relatedwork}
\end{table}

% Commercial components
\emph{LARIAT}~\cite{rossey2002lincoln} extends the testbed of the well-known DARPA 1998 and 1999 intrusion detection evaluations~\cite{lippmann2000analysis} and offers sophisticated adversary and user emulation.
However, a proprietary traffic generator and its use of physical machines prohibit the free generation of datasets with commodity hardware by others.
Likewise, the \emph{National Cyber Range}~\cite{ferguson2014national}, an effort of DARPA to build a large-scale, diverse physical testbed for cybersecurity testing, consists of only one instance and access is restricted.
\emph{CyberVAN}~\cite{chadha2016cybervan} is a complex testbed including user emulation, but relies on commercial components and is only accessible to selected entities.
Since attacks are performed manually, the testbed also lacks determinism.

% No Similarities
\emph{ViSe}~\cite{richmond2005vise} is a testbed based on virtual machines (VMs) and focuses on the forensic analysis of exploits against common operating systems.
It lacks multi-step attacks and user emulation, thus providing only limited realism.
\emph{DETERLab}~\cite{benzel2011science} is a cybersecurity testbed provided as a web service.
Registered researchers can create and run experiments remotely with choosable topology, nodes, and user/adversary emulation.
The code for the testbed is not publicly available, it does not run on commodity hardware, and it does not provide self-tests.

% Focus on Windows Machines / Azure
MITRE \emph{ATT\&CK Evaluations}~\cite{mitre2020attackevals} assess cybersecurity products on a yearly basis utilizing a network environment that consists of Microsoft Windows VMs provided by Microsoft Azure.
Rough information on how the environment was build is publicly available, but not sufficiently detailed to completely reproduce the performed experiments.
Furthermore, a lack of user emulation impairs realism.
\emph{DetectionLab}~\cite{long2021detection} is a testbed for Windows domain logging, focusing on a quick setup with security tooling and best-practice logging.
It does not implement a benign user emulation and also lacks deterministic activity.
\emph{SimuLand}~\cite{microsoft2021simuland} is a recent open-source approach for deploying lab environments with attacks and detection mechanisms in place.
Although deployment instructions and scripts are freely available, a commercial license is needed to deploy the labs to Azure.
Furthermore, there is no automation for benign or malicious activity, which contradicts the requirement of determinism for generated artifacts.

% Skopik / based on Skopik
\emph{Skopik et al.}~\cite{skopik2014semi} focus on realistic emulation of users interacting with an exemplary web application.
They mix generated legitimate activity with manually performed cyberattacks to attain more realistic log data.
However, the system's source code is not publicly available and no self-tests are included.
Finally, \emph{Landauer et al.}~\cite{landauer2021have} introduce the concept of a model-driven testbed generator, together with an implementation for web applications.
They do provide a dataset created using the testbed, but neither the testbed nor its generator are publicly available, and hence, our requirements are not met.
Furthermore, deterministic activity in dataset generation is briefly discussed, but not evaluated.

Our analysis shows that existing testbeds only insufficiently fulfill the requirements for artifact generation, especially w.r.t.\ the generation of log data:
While several testbeds emulate real-world scenarios and provide scripted/deterministic activity, none of them are designed with a focus on reproducibility and adaptability, i.e., none is available as open source software to other researchers, runs on commodity hardware, \emph{and} provides features to ease reproduction and extension such as infrastructure as code and self-tests. 

If the primary goal of a testbed is to generate \emph{network traffic} instead of log data, there are approaches that fulfill at least most of our criteria for artifact generation:
\emph{Handigol et al.}~\cite{handigol2012reproducible} examine the reproducibility of networking experiments with a focus on container-based emulation.
\emph{Wright et al.}~\cite{wright2010generating} discuss general requirements for reproducible realistic user emulation and present a model of users interacting with a graphical interface.
Notable \emph{simulators} with a focus on network cybersecurity and especially attack simulation include \emph{ADVISE}~\cite{lemay2011model}, \emph{NeSSi2}~\cite{grunewald2011agent}, and \emph{MASS}~\cite{moskal2014context}.
However, the high level of realism and detail required for generating meaningful cybersecurity log data artifacts paired with the inherent need of reproducibility and adaptability necessitates a testbed specifically designed for these purposes.

\section{SOCBED: Reproducible and Adaptable Log Data Generation}
\label{sec:implementation}

% Figure: SOCBED Overview
\begin{figure}
  \centering
  \includegraphics[width=\linewidth]{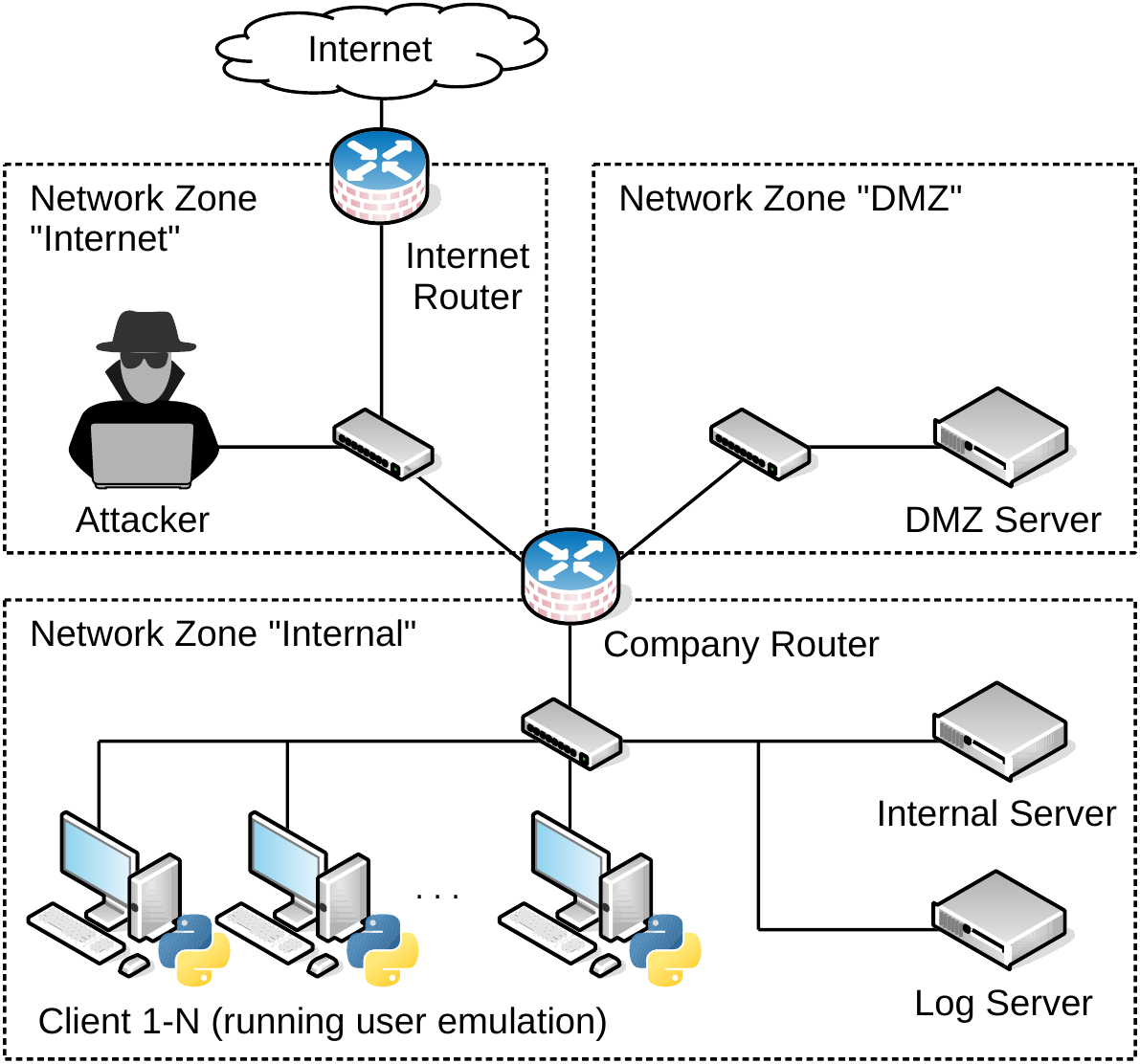}
  \caption{SOCBED emulates a small company network including benign user activity as well as multi-step cyberattacks in a reproducible and adaptable manner to facilitate sound log data generation.}
  \label{fig:overview}
  \Description{Block diagram visualizing the default systems and topology of SOCBED, as described in the text.}
\end{figure}

To address the pressing need for a testbed particularly targeting the generation of reproducible and adaptable log data as foundation for sound cybersecurity experiments, we present the design and implementation of SOCBED, our \textbf{S}elf-contained \textbf{O}pen-source \textbf{C}yberattack experimentation test\textbf{BED}.
The focus of SOCBED lies on the generation of sound log datasets for intrusion detection research, i.e., log data that are realistic, transparent, adaptable, replicable, and publicly available.
To this end, SOCBED emulates a typical company network including benign user activity and an adversary acting from the Internet or inside the company's infrastructure.
As such, SOCBED is the first proof-of-concept testbed with the goal of fostering reproducible and adaptable log data generation, thus allowing researchers to reuse or adapt existing scenarios for novel experiments to achieve better comparability of results and avoid the unnecessary effort of building new testbeds from scratch.
Most importantly, SOCBED thereby lays the foundation for long-ranging, usable log datasets fostering sound cybersecurity research by allowing frequent updates and modifications through various groups of researchers.
In the following, we describe in more detail SOCBED's emulated company network (Section~\ref{sec:vms}), infrastructure (Section~\ref{sec:infrastructure}), generation of log data (Section~\ref{sec:logdatageneration}), and measures implemented to ensure reproducibility and adaptability (Section~\ref{sec:determinism}).

\subsection{Systems and Network Topology}
\label{sec:vms}

SOCBED's network topology, resembling a typical company network, is built of seven different types of base systems, as depicted in Figure~\ref{fig:overview}.
Five of those system types mimic a small company network (Client, Internal Server, Log Server, Company Router, and DMZ Server) and are placed inside an internal network zone and a demilitarized zone (DMZ).
The other two types (Attacker and Internet Router) reside outside the company network.
Client systems can be duplicated automatically as needed and simulate benign user activity such as browsing the Internet and exchanging mails.
The Attacker system allows to run multi-step attacks that target company network systems and incorporate lateral movement, e.g., phishing a Client to visit a malicious website, escalating privileges on their machine, and exfiltrating sensible documents from another Client using that expanded access.
A dedicated Log Server system collects, processes, and stores log data from all relevant sources within the company network.
Generated log datasets can be analyzed directly on the Log Server using state-of-the-art visualization tools such as Kibana or exported for persistent offline use.

\subsection{Testbed Infrastructure}
\label{sec:infrastructure}

In general, a testbed can be realized using different types of infrastructure such as physical machines, virtual machines (VMs), or containers.
For our proof-of-concept implementation of SOCBED, we specifically decided to use VMs for the following reasons:
Physical machines are costly to operate and duplicate, which contradicts reproducibility.
Containers do not allow access to the underlying operating system for the purpose of attack execution (e.g., kernel exploits) and log data collection, hence impeding realism.
In contrast, VMs are easy to operate and duplicate, as well as able to produce realistic log data on operating system level.
Furthermore, setup and control of VMs is comparatively simple, making it possible to scale the underlying network for experimentation needs.

When building a VM-based testbed, VMs can be either self-hosted (i.e., on own hardware) or provided as a cloud service (such as Microsoft Azure).
The latter option has potential drawbacks on the reproducibility of experiments as the service provider might make changes to VMs or periphery (e.g., unavailability of older OS versions).
Additionally, such services are usually charged, contradicting our goal of reproducibility for as many researchers as possible.
According to our requirements, we chose to implement SOCBED's infrastructure using the open-source hypervisor VirtualBox.
The minimum host system requirements for basic experiments using SOCBED's base systems and network topology (cf.\ Section~\ref{sec:vms}) are 16\,GB of RAM, 30\,GB of SSD space, and a multi-core CPU with hardware-assisted virtualization.
These rather modest requirements make it possible to run SOCBED even on most modern laptops, thus enabling a majority of researchers and students to use it.

\subsection{Log Data Generation}
\label{sec:logdatageneration}

To adequately generate log data as found in a real company network under attack, SOCBED needs to implement common assets such as operating systems, services, and applications as well as an emulation of the activities of benign users and an adversary.

\paragraph{Realistic Assets}
To ensure a high degree of realism, systems in SOCBED utilize operating systems, services, and applications as typically used in company networks.
We chose to let the client systems run Windows, as it is the most common operating system for desktop and notebook computers in company networks~\cite{netapplications2019operating}.
All other systems run Linux, which is a common choice for servers and results in a smaller memory footprint of SOCBED as compared to Windows.
The servers run common services such as a web server, mail server, and domain controller.
A detailed listing of operating systems, services, and their purpose is shown in Appendix~\ref{sec:socbedservices}.

\paragraph{User Emulation}
Client systems run a user emulation to generate a ``noise floor'' of benign activity in the log data.
The user emulation is implemented as an agent running on all clients, which executes individual modules concurrently.
Its actions are logged locally.
To this end, we implemented modules for web surfing, exchanging emails, and manipulating files.
The implementation of the modules is based on configurable, seeded finite-state machines, which facilitates deterministic activity (detailed in Section~\ref{sec:determinism}).

We chose to implement web and email as they are by far the most frequent delivery methods for malware~\cite{verizon2020data}.
The web surfing module alternates periods of active web browsing sessions and prolonged periods of inactivity.
A Firefox browser window is opened at the start of each session and closed afterwards.
It is remote-controlled using the Selenium framework.
During a session, so-called \emph{routines} are executed until the session is over.
Each routine starts with either performing a web search on Google or opening a known web page directly.
The search terms and web pages are chosen from configurable lists.
When the web page has loaded, random links on the web page are followed with random delays in between.
Relevant parameters and distributions are modeled after published statistics of real users' web surfing activity~\cite{pauksztelo2014simulation}.

For the emailing module, each client has an account on the mail server, which is running on the DMZ Server VM.
The inbox is regularly checked by the module.
If a received email contains hyperlinks or attachments, they are opened automatically, facilitating phishing attacks.
% This behavior is used for some of the attack steps described earlier.
Emails are also created and sent randomly by the module.
The recipient is either another client or an external address, which replies to each incoming email with a slightly modified message.

File manipulation was implemented to increase the volume and variety of endpoint log data.
This module simulates basic file activity in a specified folder.
% By default, a local temporary folder is created.
Per iteration, a random filename and action (create, delete, append, read, move, and copy) are executed.

Supplementary custom user activity modules, e.g., videoconferencing emulation, can be added using a Python interface.
Depending on the specific use case, the implemented user activity can also be exchanged by a more sophisticated, but possibly less deterministic software such as DETERLab's DASH \cite{benzel2011science,kothari2015measuring}.

\begin{figure}
  \centering
  \includegraphics[width=\linewidth]{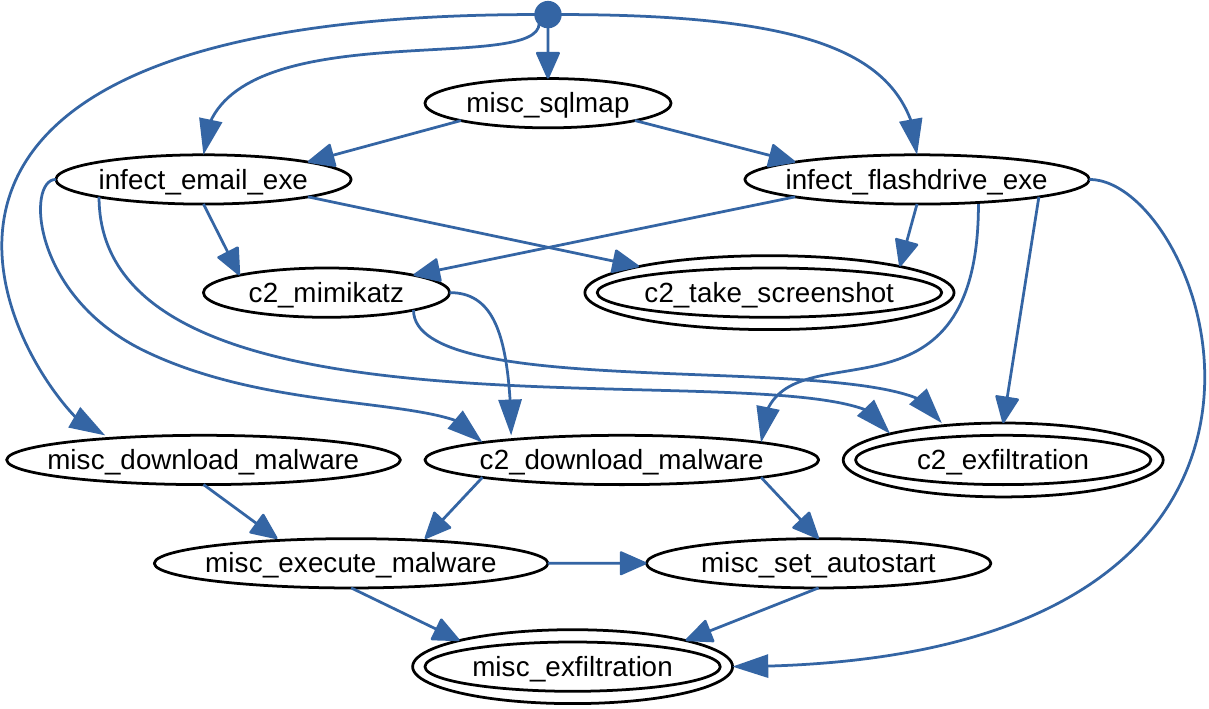}
  \caption{SOCBED can generate and execute pseudo-random multi-step cyberattacks from the implemented attack steps using a customizable digraph that models prerequisites.}
  \label{fig:graph}
  \Description{Directed graph visualizing all reasonable attack step sequences.}
\end{figure}

\paragraph{Adversary Emulation}
To replicate realistic attacks on company networks, SOCBED allows to run multi-step attacks that comprise entire kill chains~\cite{hutchins2011intelligence}, e.g., from a phishing email up to the exfiltration of confidential data.
The attack steps are implemented using common security tools (e.g., \emph{Metasploit}, \emph{mimikatz}, \emph{sqlmap}) as well as benign tools or commands often employed by adversaries (e.g., \emph{PowerShell}, \emph{xcopy}, \emph{reg}).
To choose an initial set of attack steps, we sighted publications on tactics, techniques, and procedures used by adversaries in successful network breaches, including academic research~\cite{bilge2012before,hardy2014targeted,yen2014epidemiological}, reports focusing on breach statistics~\cite{verizon2020data,symantec2015internet,mcafee2015grand}, and reports on Advanced Persistent Threat campaigns~\cite{gallagher2014wiper,kaspersky2014energetic,securelist2013redoctober} as well as the MITRE ATT\&CK Enterprise tactics and techniques~\cite{mitre2019technique}.
Currently implemented attack steps cover at least one attack step for each tactic of the ATT\&CK Matrix for Enterprise~\cite{mitre2019technique}.
A full list of implemented attack steps is shown in Appendix~\ref{sec:attackdetails}.

SOCBED is \emph{self-contained} in the sense that all cyberattacks (as well as benign activity) are fitted to the simulated company network, which is a significant advantage over stand-alone adversary emulation tools such as CALDERA~\cite{applebaum2016intelligent}, which are not fitted to a specific environment and thus are either restricted to rather simple post-exploitation steps (e.g., running local PowerShell commands) or require a lot of initial configuration to work.

As in reality, some attack steps can only execute successfully if prerequisite attack steps were executed against the same target beforehand.
In particular, there are several attack steps that use a command-and-control (C2) channel, which first has to be established by initial attack steps.
We model these dependencies with a digraph (see Figure \ref{fig:graph}) and allow to pseudo-randomly generate valid attack chains as an alternative to fixed, scripted attacks.
Analogue to the user emulation, generated attack chains solely depend on a configurable seed and are hence replicable (detailed in Section~\ref{sec:determinism}).

Cyberattacks in the real world change over time and observed campaigns reveal new attack techniques.
Therefore, SOCBED's modular adversary emulation can be adjusted to recreate different attack chains and be extended by new atomic attack steps, so-called \emph{attack modules}, using a simple Python interface.
Consequently, SOCBED can be used for both external and internal adversary models although the main focus of currently implemented modules lies on an attacker operating from the Internet.

% Figure: Logging
\begin{figure*}
  \centering
  \includegraphics[width=\textwidth]{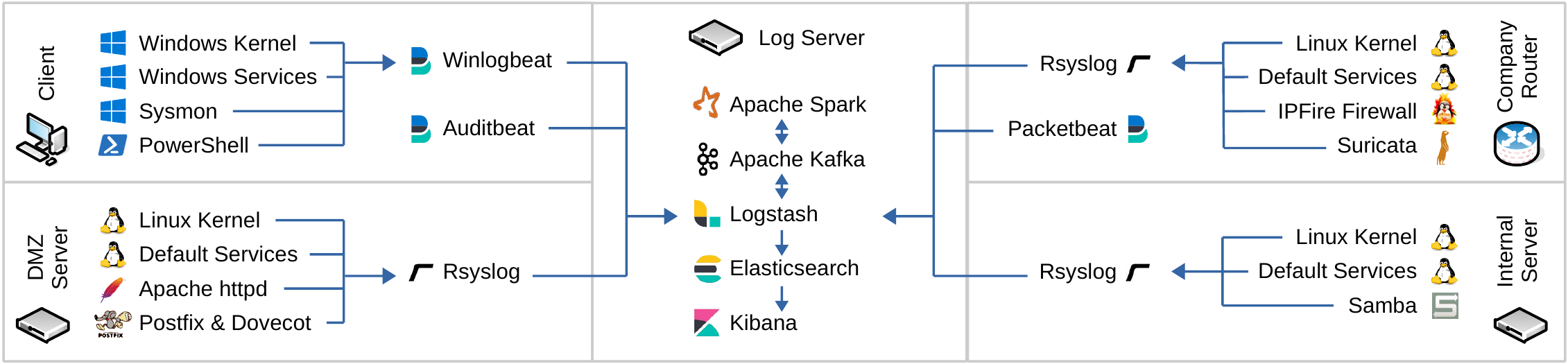}
  \caption{SOCBED collects, processes, and stores log data from best-practice sources on a dedicated log server.}
  \label{fig:logging}
  \Description{Visualization of all log data sources and their data flow from the source systems to the log server.}
\end{figure*}

\paragraph{Logging}
The main purpose of adversary and user emulation in SOCBED is to generate realistic log data that can be used for intrusion detection research.
We therefore chose log data sources and their configuration according to best practices in productive company networks~\cite{chuvakin2012logging,acsc2020windows,nist2006guide}.
Consequently, log data are collected from operating systems, services, and dedicated security software as depicted in Figure~\ref{fig:logging}.
The Windows clients can be configured to run Microsoft Sysmon and Elastic Auditbeat to capture advanced audit and security log data.
Windows Event Logs are collected and forwarded to the dedicated Log Server by an Elastic Winlogbeat agent.
The Linux machines within the simulated company network forward all syslogs (including firewall and proxy logs) to the Log Server.
The Company Router additionally runs Suricata and Packetbeat and forwards their log data as well.

The dedicated Log Server allows SOCBED users to easily search and visualize various log data from the machines in the company network.
It runs Elastic Logstash, Elasticsearch, and Kibana to collect, store, search, and visualize log data.
Additionally, log data can be exported in JSON format using the Elasticsearch API or the tool Elasticdump.
This allows to freeze generated log datasets and make them available to other researchers.

\paragraph{Network Traffic}
Although the focus of SOCBED lies on host-based log data, network traffic can be captured as well, e.g., by running the tool \emph{tcpdump} on the Company Router.
Generated and recorded traffic can then be analyzed or replayed in subsequent sessions using standard tools such as \emph{Tcpreplay}.
This setup enables strictly reproducible experiments that are based on the contents of network traffic, e.g., for NIDS evaluation.

Additionally, network flows and numerous higher-level events such as HTTP requests are captured and logged by a Packetbeat agent on the Company Router.
These logs are forwarded and stored on the log server in the same way as described for system logs above, thus enabling a straightforward analysis of network-based activity such as benign user activity or traffic caused by attacks.

\subsection{Reproducibility and Adaptability}
\label{sec:determinism}

The overarching goal of SOCBED is to showcase the feasibility of generating reproducible and adaptable log datasets.
To achieve this goal, we implemented all log-generating assets and their actions in a way that allows for deterministic activity and controlled adaptations, especially by relying on infrastructure-as-code, determinism of emulation, and self-tests as detailed in the following.

\paragraph{Infrastructure as Code}
To ensure a high level of transparency, the initial setup of all VMs is performed using infrastructure-as-code (IaC) methods.
More specifically, we use Packer and Ansible scripts to create, install, and configure all VMs automatically without user interaction.
Operating system images and additional software are automatically downloaded from the Internet and then installed and configured on the target VMs.
To avoid unintended changes in behavior, all software components are pinned to specific versions.
As SOCBED scenarios are fully defined by code, a version control system can be used to make all changes transparent and revertible, which promotes adaptability.
This approach also ensures reproducibility because different users can build the same testbed from scratch running the provided setup scripts.

\paragraph{Emulation Determinism}
Targeting determinism in emulation, we perform all activity of the user and attack emulation either scripted or pseudo-randomly with a configurable seed.
To this end, sequences of actions are generated based on finite-state machines and are logged for post-experiment investigation.
Furthermore, each client incorporates its ID into its seed, such that it behaves differently from the others but equally on each testbed run, thus making user emulation replicable.
The user emulation can retrieve websites from the Internet (for better realism) or from a web server within the simulated network (for better reproducibility).

\paragraph{Self-Tests}
SOCBED targets to realize replicable log data generation and provide easy adaptability.
Consequently, researchers using SOCBED need to be able to verify that their testbed instances are working as intended both after initial installation and after making changes.
To this end, we provide a large number of unit and system tests, which can be executed automatically using a test runner.
Unit tests check single functions for correct return values while system tests start all VMs and verify functionality of the running testbed.
More specifically, the system tests verify correct setup of VMs, execution of cyberattacks, logging, and time synchronization.
It is also possible to set up a continuous integration pipeline that rebuilds the testbed regularly (e.g., every night) and runs all tests.
Consequently, as part of our efforts for reproducibility and adaptability, our self-tests ensure correct functionality of a SOCBED setup.

Overall, by realistically reassembling a typical company network, all involved systems and assets, as well as benign user activity and adversarial actions, SOCBED provides a proof-of-concept for generating realistic log data for cybersecurity experiments.
Specifically focusing on generating reproducible and adaptable log datasets, SOCBED lays the foundation for other researchers to reproduce testbed setups on commodity computers, adapt testbed setups according to the requirements of their own research efforts, and verify the correct functionality of reproduced or adapted testbeds.

\section{Evaluation}
\label{sec:evaluation}

Sound cybersecurity experiments should be valid, controlled, and reproducible (cf.\ Section~\ref{sec:reqsci}), which imposes requirements on the used artifacts such as log data (cf.\ Section~\ref{sec:reqart}) and consequently the testbed used for generating these artifacts (cf.\ Section~\ref{sec:reqgen}).
To fulfill these requirements for generating log data artifacts and thus lay the foundation for sound cybersecurity experiments, we proposed our proof-of-concept testbed SOCBED (cf.\ Section~\ref{sec:implementation}).

In the following, we exemplarily show that it is indeed possible to perform a practical, sound experiment with log data generated by SOCBED.
We introduce the basic idea of this exemplary experiment (Section~\ref{sec:methodology}) and describe its technical setup (Sections~\ref{sec:cyberattack} and~\ref{sec:testbedsetup}).
We then present its results (Section~\ref{sec:results}) and analyze them with respect to reproducibility, controllability, and validity (Section~\ref{sec:analysis}).

\subsection{Methodology}
\label{sec:methodology}

To demonstrate SOCBED's suitability for sound cybersecurity experiments, we chose an exemplary practical experiment from the field of cyberattack detection using log data and network traffic.
More specifically, we simulate a common multi-step intrusion of an enterprise network, a topic of high practical relevance (cf.\ Section~\ref{sec:introduction}), to determine how well it can be detected with commodity detection software.
To design a concrete experiment, we narrow this research question down and formulate a hypothesis that can be tested with an experiment.

As repeatedly claimed by security experts, the default logging configuration of a modern Windows system omits numerous events that can be helpful for attack detection~\cite{acsc2020windows}.
Therefore, we decided to design an experiment to analyze whether attack detection indeed improves when switching to a best-practice logging configuration.
More precisely, our hypothesis is that when switching from the default to a best-practice configuration, more steps of an exemplary multi-step cyberattack will be detected.
By detection, we refer to at least one alert being raised as a consequence of the attack step\footnote{
In our opinion, this is a more practical metric than the total number of alerts because some attack steps yield high numbers of alerts (e.g., vulnerability scans) while others might raise only one (e.g., execution of a malicious file).}.

To test this hypothesis, we use SOCBED to recreate a small company network and launch a scripted multi-step attack against it.
We also run commodity detection software and count the true positive alerts, both with the default and best-practice logging configuration.
To prove that our experiment is reproducible, we automatically build SOCBED instances on two commodity computers, run several repetitions of the two scenarios on each of them, and then analyze the results.
To prove that the experiment is controlled, we show that changing a variable (here: the logging configuration) does not lead to unexpected side effects and thus allows to analyze the cause-effect-relationship of the change.
To show validity, we argue why the experiment results are reliable (internal validity) and can be generalized to real-world applications (external validity).

\subsection{Exemplary Multi-Step Cyberattack}
\label{sec:cyberattack}

As a concrete cyberattack, we chose a multi-step cyber espionage kill chain~\cite{hutchins2011intelligence}, as it is often executed by state-sponsored adversaries~\cite{mitre2019technique}.
We chose this type of attack because its detection is usually difficult as opposed to attacks with an obvious impact such as ransomware~\cite{friedberg2015apts}.
The attack is composed of a subset of the attack modules currently implemented in SOCBED (cf.\ Section~\ref{sec:implementation} and Appendix~\ref{sec:attackdetails}) and comprises the following steps:
(1)~An attacker probes a publicly-accessible web server of a victim company and uses SQL vulnerabilities to retrieve contact information and further details about some employees.
(2)~The attacker then sends a targeted email containing a malicious attachment to an employee.
(3)~Upon reception, the employee opens the attachment, thereby running a remote access tool that establishes a HTTP-based command-and-control~(C2) connection to the attacker.
The attacker uses the remote access tool to (4)~capture the screen of the user and (5)~retrieve cached credentials of a domain administrator using a privilege escalation.
(6)~Using these credentials, the attacker searches for another computer in the network containing interesting documents (lateral movement).
These documents are then downloaded via the C2 channel.
Finally, the attacker (7)~uploads a custom backdoor program, (8)~adds an autostart registry key, and (9)~starts the backdoor program to ensure access at a later point in time.

\subsection{Testbed Setup and Log Analysis}
\label{sec:testbedsetup}

The topology and systems for this simulation correspond to SOCBED's default setup, as depicted in Figure~\ref{fig:overview} of Section~\ref{sec:implementation}, with three client machines running.
As for detection tools, we decided to use two widespread open-source tools: Sigma rules from the official repository~\cite{github2021sigma} for log data-based detection and Suricata~\cite{oisf2021suricata} with Emerging Threat rules~\cite{proofpoint2021emergingthreatrules} for network-based detection.

We built a SOCBED instance from scratch (i.e., the infrastructure-as-code scripts created, configured, and snapshotted all virtual machines) on two notebook computers (Dell Latitude 5501 running Ubuntu 20.04 and MacBook Pro 15" Mid 2015 running macOS 10.15), each equipped with an Intel Core i7 CPU, 16\,GB of RAM, and an SSD.
For the second scenario, the Windows client was rebuilt with a best-practice logging configuration~\cite{acsc2020windows}, which mainly differs in the installation of Microsoft Sysmon~\cite{russinovich2021sysmon} and the activation of verbose PowerShell logging.
On each of the two machines, we ran ten iterations of the two scenarios described above, respectively, thus resulting in a total of 40 iterations.
Each iteration starts with booting all machines from their initial snapshots.
After 15 minutes, the described attack is launched, with three minutes idle time between the attack steps.
After 60 minutes, log data are downloaded from the machines via the Elasticsearch API, then the machines are powered off and reset to their initial state.

The downloaded log data consist of Windows Event Logs from the client machines and syslogs from the Linux machines.
For our analysis, we extracted the Suricata alerts from the syslogs and applied all suitable Sigma rules to the Windows logs.
We discarded irrelevant or false Sigma and Suricata alerts (e.g., Windows reporting usage statistics to Microsoft servers) for further analysis, thus keeping only the alerts that were caused by the attack.
Finally, we categorized these alerts by the attack step triggering them.

\subsection{Results of the Exemplary Experiment}
\label{sec:results}

\begin{table}
  \caption{We performed an exemplary experiment comprising a multi-step cyberattack ten times on two hosts. The results are consistent across all runs, thus showing that SOCBED facilitates reproducible and adaptable log data generation.}
  \small
  \centering
  \begin{tabularx}{\linewidth}{Xcccc}
    \toprule
    \multirow{2}{*}{\textbf{Attack step}}
      & \multicolumn{4}{c}{\textbf{Number of alerts}}  \\
      & \textbf{$\bar{x}_d$} & \textbf{$s_d$} & \textbf{$\bar{x}_b$} & \textbf{$s_b$} \\
    \midrule
    \multirow{2}{*}{(1) Scan and exploit web server}
      & $\mathbf{124.4}$ & $0.699$ & $\mathbf{124.1}$ & $\mathbf{1.595}$ \\
      & \cellcolor{Gray}$\mathbf{124.6}$ & \cellcolor{Gray}$0.699$ & \cellcolor{Gray}$\mathbf{124.4}$ & \cellcolor{Gray}$\mathbf{0.516}$ \\
    \multirow{2}{*}{(2) Send email with malware}
      & $2$ & $0$ & $2$ & $0$ \\
      & \cellcolor{Gray}$2$ & \cellcolor{Gray}$0$ & \cellcolor{Gray}$2$ & \cellcolor{Gray}$0$ \\
    \multirow{2}{*}{(3) Open malicious attachment}
      & $\mathbf{5.7}$ & $\mathbf{0.483}$ & $\mathbf{5.7}$ & $\mathbf{0.483}$ \\
      & \cellcolor{Gray}$\mathbf{5.9}$ & \cellcolor{Gray}$\mathbf{0.316}$ & \cellcolor{Gray}$\mathbf{5.9}$ & \cellcolor{Gray}$\mathbf{0.316}$ \\
    \multirow{2}{*}{(4) Capture screen}
      & $0$ & $0$ & $0$ & $0$ \\
      & \cellcolor{Gray}$0$ & \cellcolor{Gray}$0$ &\cellcolor{Gray}$0$ & \cellcolor{Gray}$0$ \\
    \multirow{2}{*}{(5) Collect cached credentials}
      & $0$ & $0$ & $1$ & $0$ \\
      & \cellcolor{Gray}$0$ & \cellcolor{Gray}$0$ & \cellcolor{Gray}$1$ & \cellcolor{Gray}$0$ \\
    \multirow{2}{*}{(6) Search network \& download files}
      & $0$ & $0$ & $0$ & $0$ \\
      & \cellcolor{Gray}$0$ & \cellcolor{Gray}$0$ & \cellcolor{Gray}$0$ & \cellcolor{Gray}$0$ \\
    \multirow{2}{*}{(7) Download custom backdoor}
      & $3$ & $0$ & $7$ & $0$ \\
      & \cellcolor{Gray}$3$ & \cellcolor{Gray}$0$ & \cellcolor{Gray}$7$ & \cellcolor{Gray}$0$ \\
    \multirow{2}{*}{(8) Set autostart for backdoor}
      & $0$ & $0$ & $2$ & $0$ \\
      & \cellcolor{Gray}$0$ & \cellcolor{Gray}$0$ & \cellcolor{Gray}$2$ & \cellcolor{Gray}$0$ \\
    \multirow{2}{*}{(9) Execute backdoor}
      & $0$ & $0$ & $0$ & $0$ \\
      & \cellcolor{Gray}$0$ & \cellcolor{Gray}$0$ & \cellcolor{Gray}$0$ & \cellcolor{Gray}$0$ \\
    \midrule
    \multirow{2}{*}{\textbf{Number of detected attack steps}}
      & $4$  & $0$ & $6$ & $0$ \\
      & \cellcolor{Gray}$4$  & \cellcolor{Gray}$0$ & \cellcolor{Gray}$6$ & \cellcolor{Gray}$0$ \\
    \bottomrule
  \end{tabularx}
  \label{tab:results}
\end{table}

The goal of our exemplary experiment was to test the hypothesis that the number of detected attack steps is higher when the best-practice logging configuration is used (as compared to the default configuration).
Table~\ref{tab:results} shows the sample means and standard deviations of the true positive alerts and the number of detected attack steps over all iterations ($n=10$) on both hosts for the default ($\bar{x}_d$, $s_d$) and best-practice ($\bar{x}_b$, $s_b$) configuration (Host 2 in gray, differences between the hosts in boldface).
For brevity, we pooled Suricata and Sigma alerts.
The detailed results are shown in Appendix~\ref{sec:evaldetails}.

We can see that four attack steps were detected in all iterations with the default configuration and six in all iterations with the best-practice configuration.
All standard deviations for the number of detected attack steps are zero, so there is no evidence to reject our hypothesis (the deviations in the \emph{number of alerts} are discussed in Section~\ref{sec:variations}).
We can therefore accept our hypothesis and conclude that indeed more attack steps are detected with the best-practice configuration as compared to the default configuration.

However, this does not necessarily imply causality:
The higher number of alerts could be caused by unintended side effects of the configuration change, i.e., \emph{uncontrolled} behavior.
The experiment could also have fundamental design flaws, which might be discovered by other researchers when \emph{reproducing} the experiment.
Furthermore, the results are not necessarily \emph{valid} for real-world use cases.
These potential concerns illustrate the importance of an experiment to be valid, controlled, and reproducible.

\subsection{Soundness of the Experiment}
\label{sec:analysis}

SOCBED was specifically designed for the generation of sound artifacts for log data research.
Here, we discuss how its properties support this task and thus ultimately help to make our exemplary experiment reproducible, controlled, and valid.

\paragraph{Reproducibility}
We have shown that the experiment can be performed on different machines and still leads to the same outcome, i.e., accepting the initial hypothesis.
Furthermore, the same experiment can easily be performed by other researchers because SOCBED is available as open-source software and runs on commodity hardware.
There are also no confidentiality or privacy restrictions concerning the log dataset, so it can be freely used as well.
Thus, we conclude that the experiment is indeed reproducible.

However, this does not imply that each iteration of our experiment (and thus SOCBED) produces the exact same log data (highlighted by the differences for attack steps (1) and (3) in Table~\ref{tab:results}).
Such differences result from an inherent trade-off between realism and replicability when using virtual machines for log data generation and can be attributed to different effects such as background processes and time-dependent tasks~\cite{landauer2021have,skopik2014semi}. % NOTE: analog auch in 6.6 (tex-line 639)
We further analyze the impact of such variations in Section~\ref{sec:variations} and discuss resulting limitations in Section~\ref{sec:discussion}.
The important message here is that reproducible experiments need to be designed such that they are robust against intra- and inter-host variations (just as in productive networks).

\paragraph{Controllability}
Our experiment has only one variable that is intentionally changed between runs: the Windows logging configuration.
SOCBED's infrastructure-as-code setup allows for transparent configuration changes and ensures that there are no further unintentional changes.
Built-in self-tests additionally help to verify that the functionality is not impaired by a change.
Furthermore, automated runs ensure deterministic user and adversary activity.

To confirm that the experiment is truly controlled with respect to the configuration change, we analyzed all alerts in detail.
We verified that (1) the alert types raised by the best-practice runs are a superset of the alerts with the default runs and (2) the additional best-practice alerts were truly caused by the configuration change.
Both can be easily verified, as the default configuration yields no Sigma alerts at all, which is expected as Sigma heavily builds on Sysmon as an event source.
The Suricata alerts, on the other hand, are not affected by the configuration change.
We provide more details on the types and numbers of alerts in Appendix~\ref{sec:evaldetails}.

\paragraph{Validity}
Due to the transparent infrastructure-as-code build, deterministic activity, and implemented self-tests, we have a high confidence that the testbed behaves as expected.
This is confirmed by the steady results over the different iterations.
We thus have no indication of a flawed \emph{internal} validity of the experiment.

A more difficult question to answer is whether the experiment is \emph{externally} valid, i.e., if its conclusion can be generalized and transferred to the real world.
We have insights into large enterprise networks with tens of thousands of systems that utilize the same components as our scenario (Windows 10, Sysmon, Sigma, and Suricata).
Even though we could not perform the exact same experiment in a productive network due to the risk that comes with the involved vulnerability scanning and malware execution, we are still convinced that it is indeed externally valid because all entities (operating systems, detection systems) are commonly used in practice and would thus generate similar alerts.
However, this does not necessarily mean that performing our exemplary attack against such a network would yield the \emph{exact} same alerts, e.g., due to different versions of detection rules.
Nevertheless, the experiment can serve as an indicator as to whether the analyzed configuration change should be considered in an enterprise network.

Finally, we would like to note that external validity is not a property of a testbed per se, as every testbed fails to recreate at least some properties of real-world productive systems.
Instead, external validity has to be considered for each specific experiment performed using a certain testbed.
Consequently, we can also think of experiments performed using SOCBED that would likely yield invalid results, e.g., evaluating anomaly detection methods that require a huge variety in user activity to function properly. 
However, a notable advantage of SOCBED over fixed datasets is that researchers can adapt individual parts of the testbed to make their experiments valid.

\begin{figure}
  \centering
  \includegraphics[width=\linewidth]{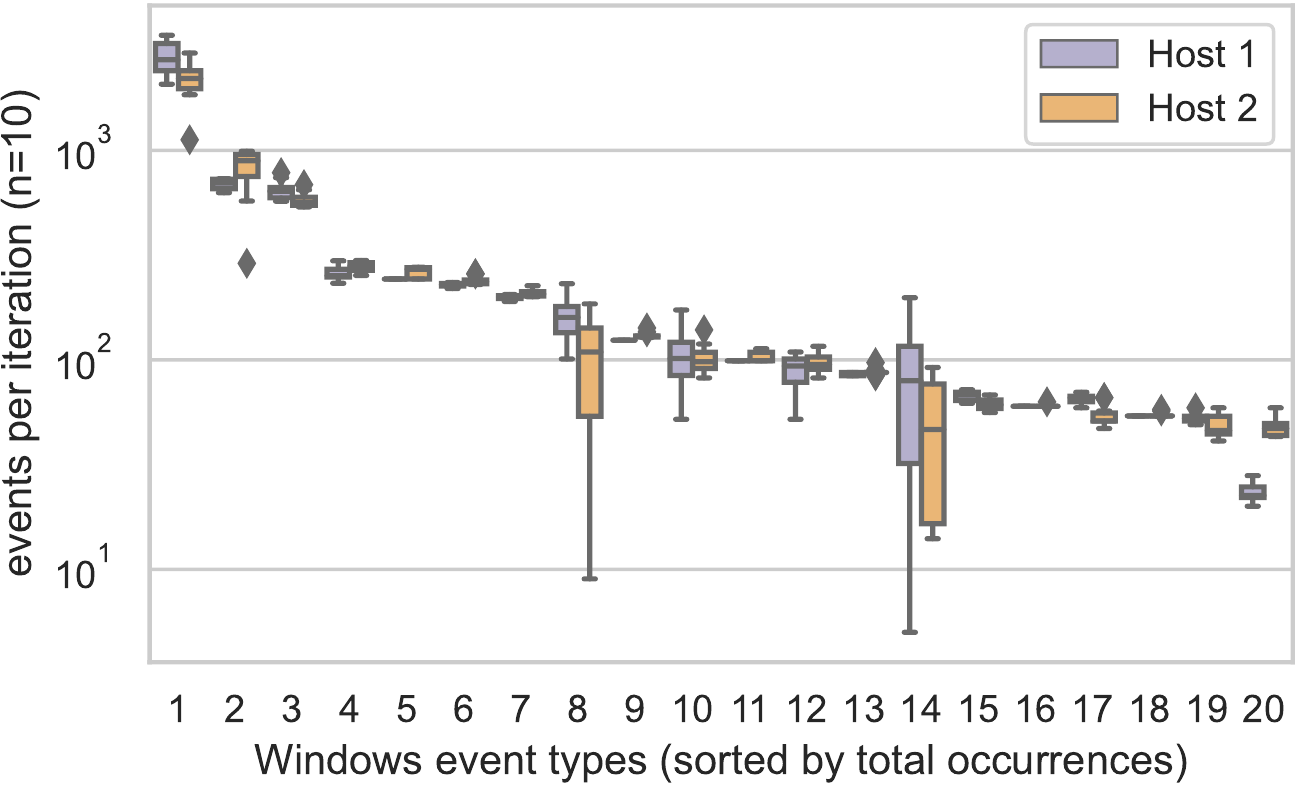}
  \caption{In contrast to the consistent log data caused by the attacks, the ``noise floor'' of operating system logs shows notable differences between the runs, even though VMs are started from the same snapshots and all activity is scripted. This is common operating system behavior and must be considered when designing experiments.}
  \label{fig:events}
  \Description{Box plot showing that some of the 20 most frequent Windows event types significantly differ in their number of occurrence between runs and hosts.}
\end{figure}

\subsection{Deep Dive: Variations in Log Data}
\label{sec:variations}

While we have shown that our exemplary experiment is reproducible, this does not imply that each iteration of SOCBED generates the exact same set of log data.
Various factors such as background processes and time-dependent tasks influence the generated log data, resulting in variations between iterations and, possibly, also host computers~\cite{landauer2021have,skopik2014semi}.
This is an unavoidable trade-off between realism and replicability when generating log data using real systems.
In the following, we analyze variations in the log data generated during our experiment in more detail and particularly quantify differences between the two hosts.
To the best of our knowledge, this important issue has never been examined before.

Table~\ref{tab:results} shows that attack steps (1) and (3) exhibit variations in the number of alerts.
These variations are caused by a different number of occurrence of two Suricata alerts (cf.\ Appendix~\ref{sec:evaldetails}).
The first one is raised during the SQL injection attack and the second one shortly after execution of the email attachment, when the shellcode is downloaded~\cite{moore2011meterpreter}.
We were able to manually reproduce these differences on both hosts, so they indeed occur and are not a flaw of the analysis process.
Still, we wanted to check whether the mean number of alerts differs between the two hosts.
To do this, we performed a two-tailed two-sample unpaired Welch's t-test~\cite{szczepanek2020ttest,mcdonald2014statistics} for the two differing alerts, which required us to run additional iterations of the triggering attack steps to obtain significant ($\alpha=0.05$) results.
Indeed, we found that for both alerts, we had to reject the test's null hypothesis that the mean number is equal on both hosts (p-values $0.000392$ and $0.00251$, respectively).
We suppose that the variations were caused by slight performance differences between iterations and hosts that resulted in a different rate of dropped packets (Suricata reported drop rates of 0.001-0.3\,\%).

Furthermore, we also took a closer look at the log data ``noise floor'', i.e., data that are not (primarily) caused by the attacks.
To give an idea of these variations, we analyzed the number of occurrences of each Windows event type (defined by the combination of provider name and  ID) over all iterations.
Figure~\ref{fig:events} shows the 20 most frequent of the 138 total event types for the default configuration on the two hosts (legend in Appendix~\ref{sec:evaldetails}).
The plot for the best practice configuration (not shown) looks similar except for several Sysmon events in the top 20 types (172 types in total).
We can see significant variations between runs for certain event types, but no striking differences between the two hosts.

In conclusion, our analysis of the Suricata alerts and Windows Event Log types shows that statistical variations between iterations and hosts indeed occur and should be anticipated.
Experiment designers should keep this in mind and perform evaluations that are robust to such intra- and inter-host variations.
This fact once again emphasizes the importance of soundly controlled and reproducible experiments so that variations caused by uncontrolled variables or non-deterministic activity can be ruled out in the analysis.

Summarizing the whole evaluation, our exemplary experiment of detecting a common multi-step intrusion of an enterprise network has shown that it is indeed possible to perform valid, controlled, and reproducible cybersecurity experiments based on log data generated with SOCBED, thus fostering research that can be built upon.

\section{Discussion and Limitations}
\label{sec:discussion}

In this paper, we have proposed SOCBED, an open-source, virtual machine (VM)-based testbed designed with reproducibility and adaptability in mind that addresses several problems of current approaches and enables researchers to conduct sound experiments.
However, every design decision also comes with potential limitations.
In the following, based on our experiences while designing, implementing, and evaluating SOCBED, we share lessons learned, discuss trade-offs, and identify further use cases for SOCBED.

To begin with, \emph{emulating a real-world scenario} may imply trade-offs with regard to reproducibility.
For example, most operating systems regularly check for updates and some even download them automatically, thus making log data and network traffic depend on the time of day and the availability of updates.
If the reproducibility of an experiment is impeded by such variations, Internet access should be disabled (but otherwise enabled for better realism).

Another requirement for reproducibility is the \emph{execution of deterministic activity}.
Yet, some experiments might comprise activity of a human adversary or user and depend on their exact timing (e.g., for anomaly detection).
In this case, we recommend to record the activity and replay it using a script to ensure reproducibility.
Likewise, for certain experiments a strong degree of determinism in network activity might be required, e.g., for an evaluation of time-based SQL attacks.
To this end, SOCBED already provides the infrastructure to retrieve websites from within the simulated network to strengthen reproducibility (cf.\ Section~\ref{sec:determinism}).
If an experiment requires an even higher degree of determinism in network activity, SOCBED's modular approach allows to extend it with a man-in-the-middle proxy (e.g., \emph{mitmproxy}~\cite{github2021mitmproxy}), including the capability of intercepting TLS encrypted communication, to deterministically replay previously recorded network traffic.

From a different perspective, while the \emph{open-source} infrastructure-as-code setup enables complete transparency and adaptability of a testbed, it comes with the challenge of occasionally disappearing software download links.
We experienced a few cases where our automatic daily builds of SOCBED broke because software repositories or URLs for downloading operating system images changed and had to be updated in the SOCBED code.
This might especially be an issue when reproducing testbed versions that are several years old.
We therefore recommend to keep local copies of all downloaded software and/or VM images if updating versions could impede the conducted experiments.

Another trade-off of SOCBED results from the fact that it runs on \emph{commodity hardware} and uses virtual machines.
In contrast to simulations, VM-based testbeds run in realtime and may behave slightly different depending on the host's soft- and hardware (just as physical systems).
Our evaluation showed that even similar hosts may lead to slight variations in generated log data.
We thus suggest to avoid running a testbed on hosts with scarce resources or background activity and to closely monitor indicators of performance issues during experiments to avoid uncontrolled behavior.

From a similar perspective, SOCBED focuses on scenarios with bounded \emph{scalability} requirements to be able to provide a high level of detail when emulating systems (i.e., full OS emulation), as required for realistic log generation.
Here, SOCBED's scalability is primarily influenced by the number of virtual machines, not by the complexity of the underlying network topology.
Given this design trade-off, very large-scale simulations requiring less realistic emulation but striving for complex scenarios with thousands of systems are out of scope for SOCBED.
In such scenarios, approaches producing large amounts of fake log data (e.g., \emph{flog}~\cite{mingrammer2020flog}) might be a better fit than SOCBED.
For the scope of this paper, we deliberately chose a small scenario with only few emulated systems which can be executed on commodity hardware to ease reproducibility.
Still, outside the scope of this paper, we successfully scaled SOCBED to execute experiments with more than a hundred realistically emulated systems using a proprietary hypervisor running on dedicated hardware (VMware ESXi).

Finally, built-in \emph{self-tests} demand additional effort during development.
Yet, we found them to be extremely valuable for discovering errors when installing SOCBED on a new host or adapting it on an existing host.
Tests are an established best practice in software development~\cite{martin2009clean} and we also strongly recommend to use and maintain them when using SOCBED or developing other testbeds.

\section{Conclusion}
\label{sec:conclusion}

Various fields of cybersecurity research base their evaluations on artifacts (e.g., log data or network traffic) that are either not publicly available or are generated using proprietary testbeds, thus heavily restricting reproducibility of their findings.
Furthermore, other researchers struggle to build on existing work because they cannot adapt existing artifact datasets for their purposes, e.g, by re-running a scenario with different attacks, other software versions, or a changed logging configuration.
Likewise, fixed datasets can lead to invalid conclusions as researchers using them might not be able to assess the appropriateness of a dataset for their own use cases due to a lack of transparency in artifact generation.

To address this issue, in this work, we started by deriving requirements for generating artifacts for cybersecurity experiments that are realistic, transparent, adaptable, replicable, and publicly available.
Based on these requirements, we argued that artifact generation for scientific experiments should be performed with testbeds that are specifically designed with a focus on reproducibility and adaptability.
As a proof-of-concept implementation, we presented SOCBED, an open-source testbed specifically targeting the generation of realistic log data for cybersecurity experiments that runs on commodity hardware.
To the best of our knowledge, SOCBED is the first testbed for log data generation that is specifically designed to foster reproducibility and adaptability, which is achieved through measures such as infrastructure as code, deterministic activity, and comprehensive self-tests.

To evaluate the reproducibility and adaptability of log data generated by SOCBED, we performed an exemplary, practical experiment from the domain of intrusion detection and showed that, even though log data naturally exhibit some variation between runs, the experiment itself is reproducible on different computers and adaptations can be performed in a controlled way.
We make the evaluation scripts and generated log dataset publicly available~\cite{uetz2021socbed,uetz2021dataset}, thus enabling others to fully reproduce our experiment.

In conclusion, our work paves the way for better reproducibility in cybersecurity research, especially in the area of log data and intrusion detection research, and consequently increases the potential to build future research efforts on existing work.

\begin{acks}
We would like to thank the anonymous reviewers and our shepherd Evangelos Markatos for their valuable feedback and fruitful comments.
This work was supported by the German Federal Ministry of Education and Research (BMBF) under grant no. 16KIS0342. The authors of this paper are responsible for its content.
\end{acks}

\begin{appendix}

\section{SOCBED Systems and Services}
\label{sec:socbedservices}

\begin{table}
  \renewcommand\thetable{6}
  \caption{Top 20 (by occurrence) Windows event types as shown in Figure~\ref{fig:events} with their mean number of occurrences over 20 iterations using the default logging configuration.}
  \small
  \centering
  \begin{tabularx}{\linewidth}{rXcr}
    \toprule
    \textbf{\#} & \textbf{Provider name} & \textbf{ID} & \multicolumn{1}{c}{$\mathbf{\bar{x}}$} \\
    \midrule
     1 & Microsoft-Windows-Security-Auditing & 5379 & 4928.7 \\
     2 & Microsoft-Windows-Security-Auditing & 5061 & 1499.8 \\
     3 & Microsoft-Windows-WindowsUpdateClient & 44 & 1231.0 \\
     4 & Microsoft-Windows-Kernel-General & 16 & 537.0 \\
     5 & PowerShell & 600 & 504.6 \\
     6 & Microsoft-Windows-Security-Auditing & 4624 & 466.5 \\
     7 & Microsoft-Windows-Security-Auditing & 4672 & 407.7 \\
     8 & Microsoft-Windows-DistributedCOM & 10010 & 257.1 \\
     9 & Microsoft-Windows-Security-Auditing & 4799 & 255.2 \\
    10 & Microsoft-Windows-WindowsUpdateClient & 19 & 206.6 \\
    11 & Microsoft-Windows-Security-Auditing & 4688 & 202.2 \\
    12 & Microsoft-Windows-WindowsUpdateClient & 43 & 185.1 \\
    13 & Microsoft-Windows-FilterManager & 6 & 173.2 \\
    14 & Windows Error Reporting & 1001 & 137.2 \\
    15 & ESENT & 642 & 128.8 \\
    16 & Microsoft-Windows-Security-Auditing & 4798 & 120.3 \\
    17 & Microsoft-Windows-Security-Auditing & 5058 & 119.3 \\
    18 & Microsoft-Windows-Security-Auditing & 4648 & 108.7 \\
    19 & Microsoft-Windows-Security-SPP & 1003 & 101.1 \\
    20 & Microsoft-Windows-Security-SPP & 16394 & 71.0 \\
    \bottomrule
  \end{tabularx}
  \label{tab:eventtypes}
\end{table}

Table~\ref{tab:services} shows the operating systems, services, and applications running on the SOCBED base virtual machines.
Default services such as DHCP are omitted for brevity.
Most of the services and applications are required for the implemented attack steps.
Additional services can easily be added and configured using Ansible scripts.

% Table: Services
\begin{table*}
  \renewcommand\thetable{3}
  \vspace{1.5cm}
  \caption{Each SOCBED VM runs services/applications required for emulating users, adversaries, or basic network functionality.}
  \centering
  \begin{tabularx}{\linewidth}{lllX}
    \toprule
    \textbf{Virtual Machine} & \textbf{Operating system} & \textbf{Service/Application} & \textbf{Purpose} \\
    \midrule
    Attacker & Kali & Apache HTTP Server & serves malicious website \\
      & & Email handler & responds to emails \\
      & & Metasploit console & launches cyberattacks \\
      & & Meterpreter HTTP listener & accepts connections \\
    \rowcolor{Gray}
    Client & Windows & Firefox & retrieves web pages \\
      \rowcolor{Gray}
      &  & User Emulation & generates user activity, opens email attachments and links \\
    Company Router & IPFire & NTP server & synchronizes time \\
      & & Squid & provides HTTP proxy \\
    \rowcolor{Gray}
    DMZ Server & Ubuntu & Damn Vulnerable Web App & gets exploited \\
      \rowcolor{Gray}
      &  & Postfix \& Dovecot & transfers and delivers emails \\
    Internal Server & Ubuntu & Samba & acts as Windows Domain Controller \\
    \rowcolor{Gray}
    Internet Router & IPFire & NTP server & synchronizes time \\
      \rowcolor{Gray}
      &  & Squid & provides HTTP proxy \\
    Log Server & Ubuntu & Elasticsearch & stores log data \\
      & & Logstash & collects log data \\
      & & Kibana & searches and visualizes log data \\
    \bottomrule
  \end{tabularx}
  \label{tab:services}
\end{table*}

\section{SOCBED Attack Step Details}
\label{sec:attackdetails}

\begin{table*}
  \renewcommand\thetable{4}
  \vspace{2.0cm}
  \caption{The attack steps currently implemented in SOCBED cover all ATT\&CK tactics (filled circles indicate coverage).}
  \centering
  \setlength\tabcolsep{3pt}
  \begin{tabularx}{\linewidth}{lXcccccccccccccc}
    \toprule
    \textbf{Attack name} &
      \textbf{Description} &
      \rotatebox[origin=l]{90}{\textbf{Reconnaissance}} &
      \rotatebox[origin=l]{90}{\textbf{\hspace{0.3mm}Resource Development\hspace{2mm}}} &
      \rotatebox[origin=l]{90}{\textbf{Initial Access}} &
      \rotatebox[origin=l]{90}{\textbf{\hspace{0.1mm}Execution}} &
      \rotatebox[origin=l]{90}{\textbf{\hspace{0.1mm}Persistence}} &
      \rotatebox[origin=l]{90}{\textbf{\hspace{0.4mm}Privilege Escalation}} &
      \rotatebox[origin=l]{90}{\textbf{Defense Evasion}} &
      \rotatebox[origin=l]{90}{\textbf{Credential Access}} &
      \rotatebox[origin=l]{90}{\textbf{\hspace{0.4mm}Discovery}} &
      \rotatebox[origin=l]{90}{\textbf{Lateral Movement}} &
      \rotatebox[origin=l]{90}{\textbf{Collection}} &
      \rotatebox[origin=l]{90}{\textbf{Command and Control}} &
      \rotatebox[origin=l]{90}{\textbf{Exfiltration}} &
      \rotatebox[origin=l]{90}{\textbf{\hspace{0.5mm}Impact}} \\
    \midrule
    \texttt{infect\_email\_exe} & Sends an email containing an infected executable file &
      \includegraphics{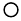} &
      \includegraphics{circle-empty-small.pdf} &
      \includegraphics{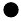} &
      \includegraphics{circle-full-small.pdf} &
      \includegraphics{circle-empty-small.pdf} &
      \includegraphics{circle-empty-small.pdf} &
      \includegraphics{circle-empty-small.pdf} &
      \includegraphics{circle-empty-small.pdf} &
      \includegraphics{circle-empty-small.pdf} &
      \includegraphics{circle-empty-small.pdf} &
      \includegraphics{circle-empty-small.pdf} &
      \includegraphics{circle-empty-small.pdf} &
      \includegraphics{circle-empty-small.pdf} &
      \includegraphics{circle-empty-small.pdf} \\
    \rowcolor{Gray}
    \texttt{infect\_flashdrive\_exe} & Mounts a drive and runs an infected exe file &
      \includegraphics{circle-empty-small.pdf} &
      \includegraphics{circle-empty-small.pdf} &
      \includegraphics{circle-empty-small.pdf} &
      \includegraphics{circle-full-small.pdf} &
      \includegraphics{circle-empty-small.pdf} &
      \includegraphics{circle-empty-small.pdf} &
      \includegraphics{circle-empty-small.pdf} &
      \includegraphics{circle-empty-small.pdf} &
      \includegraphics{circle-empty-small.pdf} &
      \includegraphics{circle-empty-small.pdf} &
      \includegraphics{circle-empty-small.pdf} &
      \includegraphics{circle-empty-small.pdf} &
      \includegraphics{circle-empty-small.pdf} &
      \includegraphics{circle-empty-small.pdf} \\
    \texttt{c2\_change\_wallpaper} & Changes the wallpaper on the target host &
      \includegraphics{circle-empty-small.pdf} &
      \includegraphics{circle-empty-small.pdf} &
      \includegraphics{circle-empty-small.pdf} &
      \includegraphics{circle-empty-small.pdf} &
      \includegraphics{circle-empty-small.pdf} &
      \includegraphics{circle-empty-small.pdf} &
      \includegraphics{circle-empty-small.pdf} &
      \includegraphics{circle-empty-small.pdf} &
      \includegraphics{circle-empty-small.pdf} &
      \includegraphics{circle-empty-small.pdf} &
      \includegraphics{circle-empty-small.pdf} &
      \includegraphics{circle-empty-small.pdf} &
      \includegraphics{circle-empty-small.pdf} &
      \includegraphics{circle-full-small.pdf} \\
    \rowcolor{Gray}
    \texttt{c2\_download\_malware} & Downloads malware through Meterpreter &
      \includegraphics{circle-empty-small.pdf} &
      \includegraphics{circle-empty-small.pdf} &
      \includegraphics{circle-empty-small.pdf} &
      \includegraphics{circle-empty-small.pdf} &
      \includegraphics{circle-empty-small.pdf} &
      \includegraphics{circle-empty-small.pdf} &
      \includegraphics{circle-empty-small.pdf} &
      \includegraphics{circle-empty-small.pdf} &
      \includegraphics{circle-empty-small.pdf} &
      \includegraphics{circle-empty-small.pdf} &
      \includegraphics{circle-empty-small.pdf} &
      \includegraphics{circle-full-small.pdf} &
      \includegraphics{circle-empty-small.pdf} &
      \includegraphics{circle-empty-small.pdf} \\
    \texttt{c2\_exfiltration} & Finds and sends documents over the C\&C channel &
      \includegraphics{circle-empty-small.pdf} &
      \includegraphics{circle-empty-small.pdf} &
      \includegraphics{circle-empty-small.pdf} &
      \includegraphics{circle-empty-small.pdf} &
      \includegraphics{circle-empty-small.pdf} &
      \includegraphics{circle-empty-small.pdf} &
      \includegraphics{circle-empty-small.pdf} &
      \includegraphics{circle-empty-small.pdf} &
      \includegraphics{circle-empty-small.pdf} &
      \includegraphics{circle-full-small.pdf} &
      \includegraphics{circle-full-small.pdf} &
      \includegraphics{circle-full-small.pdf} &
      \includegraphics{circle-full-small.pdf} &
      \includegraphics{circle-empty-small.pdf} \\
    \rowcolor{Gray}
    \texttt{c2\_mimikatz} & Obtains cached credentials using mimikatz &
      \includegraphics{circle-empty-small.pdf} &
      \includegraphics{circle-empty-small.pdf} &
      \includegraphics{circle-empty-small.pdf} &
      \includegraphics{circle-empty-small.pdf} &
      \includegraphics{circle-empty-small.pdf} &
      \includegraphics{circle-full-small.pdf} &
      \includegraphics{circle-full-small.pdf} &
      \includegraphics{circle-full-small.pdf} &
      \includegraphics{circle-empty-small.pdf} &
      \includegraphics{circle-empty-small.pdf} &
      \includegraphics{circle-empty-small.pdf} &
      \includegraphics{circle-empty-small.pdf} &
      \includegraphics{circle-empty-small.pdf} &
      \includegraphics{circle-empty-small.pdf} \\
    \texttt{c2\_take\_screenshot} & Takes a screenshot and downloads it &
      \includegraphics{circle-empty-small.pdf} &
      \includegraphics{circle-empty-small.pdf} &
      \includegraphics{circle-empty-small.pdf} &
      \includegraphics{circle-empty-small.pdf} &
      \includegraphics{circle-empty-small.pdf} &
      \includegraphics{circle-empty-small.pdf} &
      \includegraphics{circle-empty-small.pdf} &
      \includegraphics{circle-empty-small.pdf} &
      \includegraphics{circle-empty-small.pdf} &
      \includegraphics{circle-empty-small.pdf} &
      \includegraphics{circle-full-small.pdf} &
      \includegraphics{circle-full-small.pdf} &
      \includegraphics{circle-full-small.pdf} &
      \includegraphics{circle-empty-small.pdf} \\
    \rowcolor{Gray}
    \texttt{misc\_download\_malware} & Downloads malware from a web server &
      \includegraphics{circle-empty-small.pdf} &
      \includegraphics{circle-empty-small.pdf} &
      \includegraphics{circle-empty-small.pdf} &
      \includegraphics{circle-full-small.pdf} &
      \includegraphics{circle-empty-small.pdf} &
      \includegraphics{circle-empty-small.pdf} &
      \includegraphics{circle-empty-small.pdf} &
      \includegraphics{circle-empty-small.pdf} &
      \includegraphics{circle-empty-small.pdf} &
      \includegraphics{circle-empty-small.pdf} &
      \includegraphics{circle-empty-small.pdf} &
      \includegraphics{circle-empty-small.pdf} &
      \includegraphics{circle-empty-small.pdf} &
      \includegraphics{circle-empty-small.pdf} \\
    \texttt{misc\_execute\_malware} & Executes a malicious binary &
      \includegraphics{circle-empty-small.pdf} &
      \includegraphics{circle-empty-small.pdf} &
      \includegraphics{circle-empty-small.pdf} &
      \includegraphics{circle-full-small.pdf} &
      \includegraphics{circle-empty-small.pdf} &
      \includegraphics{circle-empty-small.pdf} &
      \includegraphics{circle-empty-small.pdf} &
      \includegraphics{circle-empty-small.pdf} &
      \includegraphics{circle-empty-small.pdf} &
      \includegraphics{circle-empty-small.pdf} &
      \includegraphics{circle-empty-small.pdf} &
      \includegraphics{circle-empty-small.pdf} &
      \includegraphics{circle-empty-small.pdf} &
      \includegraphics{circle-empty-small.pdf} \\
    \rowcolor{Gray}
    \texttt{misc\_exfiltration} & Copies files to a removable drive &
      \includegraphics{circle-empty-small.pdf} &
      \includegraphics{circle-empty-small.pdf} &
      \includegraphics{circle-empty-small.pdf} &
      \includegraphics{circle-empty-small.pdf} &
      \includegraphics{circle-empty-small.pdf} &
      \includegraphics{circle-empty-small.pdf} &
      \includegraphics{circle-empty-small.pdf} &
      \includegraphics{circle-empty-small.pdf} &
      \includegraphics{circle-full-small.pdf} &
      \includegraphics{circle-empty-small.pdf} &
      \includegraphics{circle-empty-small.pdf} &
      \includegraphics{circle-full-small.pdf} &
      \includegraphics{circle-empty-small.pdf} &
      \includegraphics{circle-empty-small.pdf} \\
    \texttt{misc\_set\_autostart} & Sets an autostart in the Windows registry &
      \includegraphics{circle-empty-small.pdf} &
      \includegraphics{circle-empty-small.pdf} &
      \includegraphics{circle-empty-small.pdf} &
      \includegraphics{circle-empty-small.pdf} &
      \includegraphics{circle-full-small.pdf} &
      \includegraphics{circle-empty-small.pdf} &
      \includegraphics{circle-empty-small.pdf} &
      \includegraphics{circle-empty-small.pdf} &
      \includegraphics{circle-empty-small.pdf} &
      \includegraphics{circle-empty-small.pdf} &
      \includegraphics{circle-empty-small.pdf} &
      \includegraphics{circle-empty-small.pdf} &
      \includegraphics{circle-empty-small.pdf} &
      \includegraphics{circle-empty-small.pdf} \\
    \rowcolor{Gray}
    \texttt{misc\_sqlmap} & Performs an SQL injection attack to steal credentials &
      \includegraphics{circle-full-small.pdf} &
      \includegraphics{circle-full-small.pdf} &
      \includegraphics{circle-full-small.pdf} &
      \includegraphics{circle-empty-small.pdf} &
      \includegraphics{circle-empty-small.pdf} &
      \includegraphics{circle-empty-small.pdf} &
      \includegraphics{circle-empty-small.pdf} &
      \includegraphics{circle-full-small.pdf} &
      \includegraphics{circle-empty-small.pdf} &
      \includegraphics{circle-empty-small.pdf} &
      \includegraphics{circle-empty-small.pdf} &
      \includegraphics{circle-empty-small.pdf} &
      \includegraphics{circle-empty-small.pdf} &
      \includegraphics{circle-empty-small.pdf} \\
    \bottomrule
  \end{tabularx}
  \label{tab:attacks}
\end{table*}

Table~\ref{tab:attacks} shows the attack steps currently implemented in SOCBED.
There are three types of attack steps.
Steps starting with ``infect'' create initial access to the company network by exploiting the targeted client and then running a payload.
This payload establishes a command and control (C2) channel to the Attacker VM using Metasploit's \emph{Meterpreter} reverse HTTP module~\cite{moore2011meterpreter}.
Steps starting with ``c2'' use this channel to execute further actions via Meterpreter.
They can only execute successfully when an ``infect'' attack step has been executed before on the same client and the payload is still active, i.e., the client is not rebooted and the process is not killed.
Steps starting with ``misc'' are self-contained and can be executed independently.
They can mimic, e.g., an internal adversary, an employee falling victim to a social engineering attack, or an attack step triggered by an already active malware.

Some of the implemented attack steps are commonly performed by external attackers (\emph{infect\_email\_*}, \emph{c2\_*}), while others resemble internal attackers (\emph{misc\_exfiltration}, \emph{infect\_flashdrive\_exe}) or both alike (remaining \emph{misc\_*} attacks).

\section{Detailed Evaluation Results}
\label{sec:evaldetails}

\begin{table*}
  \renewcommand\thetable{5}
  \vspace{1cm}
  \caption{Alerts per rule for our exemplary scenarios ($d$efault and $b$est-practice logging configuration). The two lines per rule show the results (sample mean and SD, $n=10$) for the two hosts on which the experiment was performed, respectively (second host with gray background, differences in boldface).}
  \small
  \centering
  \begin{tabularx}{\linewidth}{cXcccc}
    \toprule
    & \textbf{Rule name} & \multicolumn{4}{c}{\textbf{Number of alerts}} \\
    &                    & \multicolumn{1}{c}{$\bar{x}_d$} & \multicolumn{1}{c}{$s_d$} & \multicolumn{1}{c}{$\bar{x}_b$} & \multicolumn{1}{c}{$s_b$} \\
    \midrule
    \parbox[t]{3mm}{\multirow{10}{*}{\rotatebox[origin=c]{90}{\textbf{Sigma}}}}
    & \multirow{2}{*}{Autorun Keys Modification}                                                                   & $0$ & $0$ & $1$ & $0$ \\
    &                                                                                                              & \cellcolor{Gray}$0$ & \cellcolor{Gray}$0$ & \cellcolor{Gray}$1$ & \cellcolor{Gray}$0$ \\
    & \multirow{2}{*}{Direct Autorun Keys Modification}                                                            & $0$ & $0$ & $1$ & $0$ \\
    &                                                                                                              & \cellcolor{Gray}$0$ & \cellcolor{Gray}$0$ & \cellcolor{Gray}$1$ & \cellcolor{Gray}$0$ \\
    & \multirow{2}{*}{Meterpreter or Cobalt Strike Getsystem Service Start}                                        & $0$ & $0$ & $1$ & $0$ \\
    &                                                                                                              & \cellcolor{Gray}$0$ & \cellcolor{Gray}$0$ & \cellcolor{Gray}$1$ & \cellcolor{Gray}$0$ \\
    & \multirow{2}{*}{Non Interactive PowerShell}                                                                  & $0$ & $0$ & $1$ & $0$ \\
    &                                                                                                              & \cellcolor{Gray}$0$ & \cellcolor{Gray}$0$ & \cellcolor{Gray}$1$ & \cellcolor{Gray}$0$ \\
    & \multirow{2}{*}{Windows PowerShell Web Request}                                                              & $0$ & $0$ & $3$ & $0$ \\
    &                                                                                                              & \cellcolor{Gray}$0$ & \cellcolor{Gray}$0$ & \cellcolor{Gray}$3$ & \cellcolor{Gray}$0$ \\
    \midrule
    \parbox[t]{3mm}{\multirow{40}{*}{\rotatebox[origin=c]{90}{\textbf{Suricata}}}}
    & \multirow{2}{*}{ET INFO EXE IsDebuggerPresent (Used in Malware Anti-Debugging)}                              & $1$ & $0$ & $1$ & $0$ \\
    &                                                                                                              & \cellcolor{Gray}$1$ & \cellcolor{Gray}$0$ & \cellcolor{Gray}$1$ & \cellcolor{Gray}$0$ \\
    & \multirow{2}{*}{ET INFO Executable Download from dotted-quad Host}                                           & $1$ & $0$ & $1$ & $0$ \\
    &                                                                                                              & \cellcolor{Gray}$1$ & \cellcolor{Gray}$0$ & \cellcolor{Gray}$1$ & \cellcolor{Gray}$0$ \\
    & \multirow{2}{*}{ET INFO Executable Retrieved With Minimal HTTP Headers - Potential Second Stage Download}    & $1$ & $0$ & $1$ & $0$ \\
    &                                                                                                              & \cellcolor{Gray}$1$ & \cellcolor{Gray}$0$ & \cellcolor{Gray}$1$ & \cellcolor{Gray}$0$ \\
    & \multirow{2}{*}{ET INFO SUSPICIOUS Dotted Quad Host MZ Response}                                             & $2$ & $0$ & $2$ & $0$ \\
    &                                                                                                              & \cellcolor{Gray}$2$ & \cellcolor{Gray}$0$ & \cellcolor{Gray}$2$ & \cellcolor{Gray}$0$ \\
    & \multirow{2}{*}{ET INFO SUSPICIOUS SMTP EXE - EXE SMTP Attachment}                                           & $2$ & $0$ & $2$ & $0$ \\
    &                                                                                                              & \cellcolor{Gray}$2$ & \cellcolor{Gray}$0$ & \cellcolor{Gray}$2$ & \cellcolor{Gray}$0$ \\
    & \multirow{2}{*}{ET POLICY PE EXE or DLL Windows file download HTTP}                                          & $2$ & $0$ & $2$ & $0$ \\
    &                                                                                                              & \cellcolor{Gray}$2$ & \cellcolor{Gray}$0$ & \cellcolor{Gray}$2$ & \cellcolor{Gray}$0$ \\
    & \multirow{2}{*}{ET SCAN Sqlmap SQL Injection Scan}                                                           & $2$ & $0$ & $2$ & $0$ \\
    &                                                                                                              & \cellcolor{Gray}$2$ & \cellcolor{Gray}$0$ & \cellcolor{Gray}$2$ & \cellcolor{Gray}$0$ \\
    & \multirow{2}{*}{ET TROJAN Possible Metasploit Payload Common Construct Bind\_API (from server)}              & $\mathbf{1.7}$ & $\mathbf{0.483}$ & $\mathbf{1.7}$ & $\mathbf{0.483}$ \\
    &                                                                                                              & \cellcolor{Gray}$\mathbf{1.9}$ & \cellcolor{Gray}$\mathbf{0.316}$ & \cellcolor{Gray}$\mathbf{1.9}$ & \cellcolor{Gray}$\mathbf{0.316}$ \\
    & \multirow{2}{*}{ET WEB\_SERVER ATTACKER SQLi - SELECT and Schema Columns}                                    & $\mathbf{6.4}$ & $0.699$ & $\mathbf{6.1}$ & $\mathbf{1.595}$ \\
    &                                                                                                              & \cellcolor{Gray}$\mathbf{6.6}$ & \cellcolor{Gray}$0.699$ & \cellcolor{Gray}$\mathbf{6.4}$ & \cellcolor{Gray}$\mathbf{0.516}$ \\
    & \multirow{2}{*}{ET WEB\_SERVER Attempt To Access MSSQL xp\_cmdshell Stored Procedure Via URI}                & $1$ & $0$ & $1$ & $0$ \\
    &                                                                                                              & \cellcolor{Gray}$1$ & \cellcolor{Gray}$0$ & \cellcolor{Gray}$1$ & \cellcolor{Gray}$0$ \\
    & \multirow{2}{*}{ET WEB\_SERVER MYSQL Benchmark Command in URI to Consume Server Resources}                   & $2$ & $0$ & $2$ & $0$ \\
    &                                                                                                              & \cellcolor{Gray}$2$ & \cellcolor{Gray}$0$ & \cellcolor{Gray}$2$ & \cellcolor{Gray}$0$ \\
    & \multirow{2}{*}{ET WEB\_SERVER MYSQL SELECT CONCAT SQL Injection Attempt}                                    & $22$ & $0$ & $22$ & $0$ \\
    &                                                                                                              & \cellcolor{Gray}$22$ & \cellcolor{Gray}$0$ & \cellcolor{Gray}$22$ & \cellcolor{Gray}$0$ \\
    & \multirow{2}{*}{ET WEB\_SERVER Possible attempt to enumerate MS SQL Server version}                          & $2$ & $0$ & $2$ & $0$ \\
    &                                                                                                              & \cellcolor{Gray}$2$ & \cellcolor{Gray}$0$ & \cellcolor{Gray}$2$ & \cellcolor{Gray}$0$ \\
    & \multirow{2}{*}{ET WEB\_SERVER Possible Attempt to Get SQL Server Version in URI using SELECT VERSION}       & $6$ & $0$ & $6$ & $0$ \\
    &                                                                                                              & \cellcolor{Gray}$6$ & \cellcolor{Gray}$0$ & \cellcolor{Gray}$6$ & \cellcolor{Gray}$0$ \\
    & \multirow{2}{*}{ET WEB\_SERVER Possible MySQL SQLi Attempt Information Schema Access}                        & $4$ & $0$ & $4$ & $0$ \\
    &                                                                                                              & \cellcolor{Gray}$4$ & \cellcolor{Gray}$0$ & \cellcolor{Gray}$4$ & \cellcolor{Gray}$0$ \\
    & \multirow{2}{*}{ET WEB\_SERVER Possible SQL Injection Attempt SELECT FROM}                                   & $16$ & $0$ & $16$ & $0$ \\
    &                                                                                                              & \cellcolor{Gray}$16$ & \cellcolor{Gray}$0$ & \cellcolor{Gray}$16$ & \cellcolor{Gray}$0$ \\
    & \multirow{2}{*}{ET WEB\_SERVER Possible SQL Injection Attempt UNION SELECT}                                  & $19$ & $0$ & $19$ & $0$ \\
    &                                                                                                              & \cellcolor{Gray}$19$ & \cellcolor{Gray}$0$ & \cellcolor{Gray}$19$ & \cellcolor{Gray}$0$ \\
    & \multirow{2}{*}{ET WEB\_SERVER Script tag in URI Possible Cross Site Scripting Attempt}                      & $1$ & $0$ & $1$ & $0$ \\
    &                                                                                                              & \cellcolor{Gray}$1$ & \cellcolor{Gray}$0$ & \cellcolor{Gray}$1$ & \cellcolor{Gray}$0$ \\
    & \multirow{2}{*}{ET WEB\_SERVER SQL Errors in HTTP 200 Response (error in your SQL syntax)}                   & $36$ & $0$ & $36$ & $0$ \\
    &                                                                                                              & \cellcolor{Gray}$36$ & \cellcolor{Gray}$0$ & \cellcolor{Gray}$36$ & \cellcolor{Gray}$0$ \\
    & \multirow{2}{*}{ET WEB\_SERVER SQL Injection Select Sleep Time Delay}                                        & $7$ & $0$ & $7$ & $0$ \\
    &                                                                                                              & \cellcolor{Gray}$7$ & \cellcolor{Gray}$0$ & \cellcolor{Gray}$7$ & \cellcolor{Gray}$0$ \\
    \bottomrule
  \end{tabularx}
  \label{tab:alertdetails}
\end{table*}

Table~\ref{tab:alertdetails} shows all Sigma and Suricata alerts that occurred during our evaluation.
For Sigma, we used all Windows Event Log-specific rules as of February 4, 2021\footnote{\url{https://github.com/SigmaHQ/sigma/tree/12054544bbac415438b2207c08bd92633a51b}}.
For Suricata, we used Emerging Threat rules as of May 4, 2021.
The latter are contained in the SOCBED repository because stale rule sets are generally not provided for download on the official website\footnote{\url{https://rules.emergingthreats.net/OPEN_download_instructions.html}}, yet a fixed version of the rules is important for the reproducibility of Suricata alerts across different SOCBED instances.

Table~\ref{tab:eventtypes} shows the top 20 (by occurrence) Windows event types as depicted in Figure~\ref{fig:events}.
All of these types commonly appear on Windows systems and are not specific to the executed attacks.
For example, the most frequent event \emph{Microsoft-Windows-Security-Auditing ID 5379} informs that ``Credential Manager credentials were read'', i.e., a user performs a read on stored credentials, e.g., during the logon process\footnote{\url{https://www.ultimatewindowssecurity.com/securitylog/encyclopedia/event.aspx?eventid=5379}}.

\end{appendix}

\end{document}